\documentclass{aa}

\usepackage{graphicx}
\usepackage{psfig}
\usepackage{natbib}
\bibpunct{(}{)}{;}{a}{}{,}
\begin{document}
\newcommand{\aap}{A\&A}
\newcommand{\aaps}{A\&AS}
\newcommand{\pasp}{PASP}
\newcommand{\apj}{ApJ}
\newcommand{\apjl}{ApJ}
\newcommand{\apjs}{ApJS}
\newcommand{\mnras}{MNRAS}
\newcommand{\aj}{AJ}
\newcommand{\araa}{ARA\&A}
\newcommand{\gca}{Geocosmochimica Acta\ }
\newcommand{\AuJP}{AuJP\ }
\newcommand{\nat}{Nat}
\newcommand{\pasj}{PASJ\ }
\newcommand{\procspie}{Proc. SPIE\ }
\newcommand{\advspres}{Adv. Sp. Res.\ }
\newcommand{\iaucirc}{IAU Circ.\ }
\newcommand{\qjras}{QJRAS\ }
\newcommand{\ch}{{\it Chandra}}
\newcommand{\xmm}{XMM-{\it Newton}}
\newcommand{\cy}{Cyg~X-2}
\newcommand{\pr}{$^{\prime}$}
\newcommand{\mg}{Mg$_2$SiO$_4$}
\newcommand{\fe}{Fe$_2$SiO$_4$}
\newcommand{\femg}{FeMgSiO$_4$}
\newcommand{\nh}{$N_{\rm H}$} 
\newcommand{\abs}{absorption}
\newcommand{\sca}{scattering}
\newcommand{\bfi}{\begin{figure}}
\newcommand{\efi}{\end{figure}}
\def\ltsim{\raisebox{-.5ex}{$\;\stackrel{<}{\sim}\;$}}
\def\gtsim{\raisebox{-.5ex}{$\;\stackrel{>}{\sim}\;$}}



\title{Absorption and scattering by interstellar dust:\\ XMM-Newton observation of \object{\cy} }

\author{E.~Costantini\inst{1,2}, M.~J.~Freyberg\inst{3} \& P.~Predehl\inst{3}}

\institute{SRON National Institute for Space Research, 
Sorbonnelaan, 2, 3584CA, Utrecht, The Netherlands
\and
Astronomical Institute, Utrecht University, P.O. Box 80000, 3508TA Utrecht, The Netherlands
\and
Max-Planck-Institut 
f\"ur extraterrestrische Physik, Giessenbachstr.\ 1, D-85748 Garching bei M\"unchen, Germany}
\date{Received / Accepted }
\authorrunning{Costantini, Freyberg \& Predehl}
\titlerunning{XMM-Newton observation of \cy }
\abstract
{
We present results of the XMM-{\it Newton} observation on the bright X-ray binary \cy. In our analysis we focus upon
 the absorption and scattering of the X-ray emission by 
interstellar dust distributed along the line of sight. 
The scattering halo around \cy, observed with the CCD 
	 detector EPIC-pn, is well detected up to $\sim$7 arcmin and contributes $\sim$5-7\% to the total source emission at 1 keV, depending on the dust size
	 distribution model considered.  
	 For the first time spatially resolved spectroscopy of a scattering halo is performed. In the halo spectrum we 
	 clearly detect the signature 
	 of the interstellar dust elements: O, Mg, and Si. In the $0.4-2$\,keV band, 
	 the spectral modeling of the halo shows a major contribution of silicates (olivine and pyroxene). 
	 The spatial analysis of the halo surface brightness profile shows that the dust 
	 is smoothly distributed toward \cy\ at least for $\sim$60\% of the path to the source. However, given the substantial pile-up, we could not 
	 investigate fainter or narrower components of the halo. 
	 Within this observation limits, the data do not show preference for a specific dust size distribution; namely 
	 the Mathis, Rumpl \& Nordsieck (1977) or the Weingartner \& Draine (2001) model. 
	 In this analysis we used the Mie theory to compute the differential scattering cross
	  section.
	The RGS data were used to investigate the ISM absorption. In particular the absorption spectrum shows 
	complexity around the oxygen edge at $\sim$0.54\,keV, 
        which cannot be explained in a unique way: absorption by molecular oxygen or ionized atomic oxygen, as proposed in other studies of \cy. 
	Combining the RGS results with the additional information on dust grains provided 
	by the EPIC-pn spectrum of the scattered radiation we estimate a column density for dust absorption by oxygen, provided that it is locked 
	in silicate grains.   
}

\maketitle

\section{Introduction}
 The observed light from a source is obscured by the interstellar matter (ISM) through the combination of two processes: absorption and scattering. 
 Absorption is due to both gas and dust, 
 whereas scattering is attributed to dust alone.    
Differently from the IR to UV wavelengths range, in the X-ray regime 
the observation of  \abs\ and \sca\ by interstellar dust (ID) are strongly coupled. Thus, in the X-ray
 regime, the two extinction mechanisms can be simultaneously observed and studied.
 If an X-ray emitter is located behind a layer of dust,
 its radiation will be absorbed  and at the same time scattered into the direction of the observer. 
In the X-rays the scattering mechanism is no longer explained by the simple Rayleigh formula. 
In particular the scattering angle is in this case very small 
($\theta_{\rm scatt}\propto (\lambda/a)\ll 1^{\circ}$), forward directed, dependent on the wavelength of the incident photon ($\lambda$)
 and the size ($a$) of the grain. 
The small scattering angle results in a halo of diffuse emission around the source \citep{1965ApJ...141..864O}. 
The energy range in which \abs\ and \sca\ can be studied, is a strong function of the equivalent hydrogen column density of the medium 
($N_{\rm H}$). Indeed the X-ray radiation is obscured by absorption depending on the value of \nh : 
$I=I_0e^{-N_{{\rm H}} \sigma}$, where $I_0$ is 
the source radiation and $\sigma$ is the
absorption cross section. 
Through the analysis of the absorbed spectrum, information on the chemistry, column density and abundances of the ID grains can be inferred.
 Simultaneously, 
the spectral and spatial properties of the X-ray halos can be analyzed. The halo intensity, angular extension, and spectral distribution 
are a function of the size distribution and composition of the scatterers (the dust grains), 
their distribution along the line of sight, and the spectral properties of the source illuminating them.\\ 
Sources with faint halos have a hydrogen column density which is low enough to not completely absorb the soft X rays. 
High sensitivity instruments are needed to study the 
emission  of the scattering halo, which is very weak compared to the brightness of a background source 
\citep[up to 20\% of the soft emission, ][]{pk96}.
Faint halos are important to study ID chemistry. Indeed the energy range where scattering occurs (approximately 
$0.3-2$\,keV), includes possible features of ID components, primarily 
 oxygen (0.54 keV), magnesium (1.3 keV), and silicon (1.84 keV) \citep[][ hereinafter D03]{pk96,2003ApJ...598.1026D}. 
An energy resolution of 
$\sim 80-150$\,eV over the energy band of interest for the scattering process, 
makes it possible to spectroscopically investigate the features of a faint halo. 
Previous studies of scattering halo profiles 
were carried out with {\it Einstein} \citep[e.g.][]{mg86,gall95} and ROSAT \citep[][ hereinafter PS95]{sd98,ps95}. 
According to those findings, the ID size distribution appeared to be consistent 
with the \citet{mrn77} (hereinafter MRN) model. The MRN model includes a mixture of 
carbonaceous and silicate 
materials, with size distribution $a^{-3.5}$, for 
$0.001<a<0.25 \mu$m. 
Alternative grain size distribution models have been proposed, differing mostly in the chosen 
boundaries of the grain size range, the slope of the distribution and the inner structure of the grain itself 
\citep[][ hereinafter WD01]{1989ApJ...341..808M,2000JGR...10510343L,wd01}. 
From ROSAT data \citep[Nova V1974 Cygni, ][]{drainetan} and the 
{\it Chandra} observation of GX~13+1 \citep{gx13} the role of grains of size much larger than $0.4 \mu$m has been stated 
not to play a major role, at least in the diffuse ISM, contrary to the Solar System environment, 
where grains up to $1\,\mu$m should have a significant contribution \citep{2001ApJ...550L.201W}. 
On the other side of the range of the grain size, 
a better understanding of the infra-red spectrum of ID allowed to state the importance of ultra-small
particles, the Aromatic Polycyclic Hydrocarbons (PAH), with radius $a<30$\,\AA. 
Such a contribution was included in the WD01 model.   
 
Finally, it has been recognized that a simple analytical computation of the differential scattering cross section, the so called Rayleigh-Gans (hereinafter RG) approximation , 
could be misleading if applied to halo energies 
$<1$\,keV and/or large grains $a>0.25 \mu$m.
The full Mie theory \citep{mie08}, from which the RG approximation is derived, had to be used \citep{sd98,drainetan}. 
The models applied to X-ray scattering halos are the result of a deeper knowledge of ID properties gathered at longer 
wavelengths. Due to the low resolution of early 
X-ray instruments, only integral
properties of the dust could be studied, adding relatively little information on the nature of ID. On the contrary, with the
 X-ray observatories now flying, we can 
address other issues like: {\it (i)} the chemical properties of dust particles that scatter X-rays, {\it (ii)} 
abundances and depletion in the ISM, 
and {\it (iii)} the actual distribution of dust along the line of sight.\\   
In this paper we present the RGS and EPIC-pn analysis of \cy, located at Galactic coordinates $l=87.33^\circ, b=-11.32^\circ$, behind a dust layer with 
equivalent $N_{\rm H}$ column density of the order of $\sim 2\times10^{21} {\rm cm}^{-2}$, which produces a relatively weak 
scattering halo. This makes \cy\ an ideal candidate to study both the spatial and spectral distribution of the halo at energies 
softer than 2 keV.  
The fractional halo intensity of \cy , defined as the intensity of the halo extended emission over the total observed emission, 
was estimated from ROSAT-PSPC to be 3.9\% at 1.06 keV (PS95). 
Now, with the high sensitivity of \xmm, the halo can be resolved and 
analyzed down to 0.4 keV with the EPIC-pn.
Absorption by the ISM toward the line of sight of \cy\ was studied with the RGS. 
Recently \cy\ was studied by \citet{tak03}, using {\it Chandra}-LETG and by \citet{juett03} using \ch -HETG. 
In each analysis, the absorption features in 
the spectral region of the oxygen edge were interpreted in different ways. 
\citet{tak03} claimed to have detected absorption by
molecular oxygen, while \citet{juett03} interpret those features in terms of mildly ionized oxygen in the ISM.

The paper is organized as follows: In Sect.~2 the principles of the scattering halo theory are presented. 
In Sect.~3 the analysis of RGS and EPIC-pn data of \cy\ 
is shown. 
Sect.~4  describes the careful
extraction of the information on scattered radiation.   
Sect.~5 is then devoted to the spatial and the spectral modeling of the scattered halo. 
Finally, in Sect.~6 we discuss our results, and in Sect.~7 the conclusions of this work are shown.


\section{The Halo Theory}
The intensity of the light scattered by dust at a scattering angle $\theta_{\rm sca}$, assuming spherical 
grains and single scattering has the following general form \citep[e.g., ][]{1991ApJ...376..490M}:
\begin{eqnarray}
I(\theta_{\rm sca})=\int_{E_{\rm min}}^{E_{\rm max}} F(E) dE\int^{a_{\rm max}}_{a_{\rm min}} n(a) da \times \nonumber \\ 
\times \int_{0}^{1} \frac{\tilde{f}(x)}{(1-x)^2} \frac{d\sigma}{d\Omega} dx. 		
\label{eq:isca}
\end{eqnarray} 

$F(E)$ is the spectral energy distribution of the source; $a$ is the grain radius with number density 
$n(a)$; $x$ is the fractional distance of the 
total path, from the source~($x=1$) to the observer~($x=0$), at which the scattering occurs; and $\tilde{f}(x)$ is the normalized spatial distribution of the scattering sites.  
Since we consider dust that is evenly distributed, $\tilde{f}(x)=1$. 
For the scattering angle $\theta_{\rm sca}$, it holds that $\theta_{\rm sca}=\theta_{\rm obs}/(1-x)$. 
The term $d\sigma/d\Omega$ is the differential scattering cross section, 
a function of $\theta$, $a$ and $E$. There are two critical terms in 
Eq~\ref{eq:isca}. One is $n(a)$, which depends on the physical and chemical state of the dust grains.
The other crucial term in evaluating the scattered emission is the differential scattering cross section for which the exact solution is given by the Mie theory 
\citep{mie08,1957lssp.book.....V}. This describes the scattering and absorption of an electromagnetic
wave by spherical solid particles.\\ 
Quantitatively, the refraction index $m$ of a given material can be written as \citep[e.g., ][]{1993ADNDT..54..181H}:\\
\begin{equation}
m=1-\frac{r_{\rm e} \lambda^2}{2\pi}\sum_q n_qf_q(0)
\end{equation}
where $r_{\rm e}$ is the classical electron radius, $n_q$ the number density of the atom $q$.
The quantity $f_q(0)$ is the complex forward atomic scattering factor for the element $q$. This definition only holds for small
scattering angles, as is the case in the X-ray band, or for wavelengths that are large compared to the electron density of the
scattering medium. 
For each element, the imaginary and real part of $f=f^{\prime}+if^{\prime\prime}$ are related to each other through the Kramers-Kronig relations \citep[e.g.][ D03]{1993ADNDT..54..181H}: 
\begin{equation}
f^{\prime}(\omega)=\frac{2}{\pi}\int^{\infty}_0 \frac{f^{\prime\prime}(\omega^\prime)\omega^\prime d\omega^\prime}{\omega^2-\omega^{\prime 2}} 
\end{equation}

\begin{equation}
f^{\prime\prime}(\omega) = \frac{\omega\mu(\omega)}{2\pi n_q r_e c} ,
\end{equation}
where $\mu(\omega)$ is the absorption coefficient at the incident energy $E=\hbar\omega$.   
Here the assumption is that for a given compound the dielectric function is the sum of the single atoms contributions.
However, at the threshold energy, the absorption coefficient strongly depends on the chemical compound. 
The interaction between a photoelectron wave and all the other waves backscattered by the neighboring atoms creates modulations in the absorption cross section. 
These are in
general called X-ray Absorption Fine Structure (XAFS). The XAFS features, which are few tenth of angstroms wide, have been recently recognized in high energy 
resolution absorption spectra of some astronomical sources \citep[e.g.,][]{2002ApJ...567.1102L}. Only instruments with energy resolution better than 
$\Delta\lambda\sim0.02$\AA\ can resolve these features. 
The differential scattering cross section, calculated using the Mie approach, as a function of energy is 
displayed in Fig.~\ref{f:mie_i} 
for various scattering angles, considering the case of 
\mg\ and a grain size of
0.1$\mu$m. The presence of $\mu(\omega)$ in the scattering cross section causes spikes at the K-edge 
of a given element, 
as seen in the figure.  
We note also that at the softer energies 
we do not expect a dramatic change in the spectral shape of the differential cross section as a function of the 
scattering angle.

In our observation, the \cy\ halo is visible down to 0.4 keV, 
as the hydrogen column density toward the source is relatively low 
\citep[$N_{\rm H}\sim2.17\times 10^{21}\,{\rm cm}^{-2}$, as measured from the H{\sc i} emission, ][]{1990ARA&A..28..215D}. 
Therefore we need to evaluate the $\frac{d\sigma}{d\Omega}$ term in 
Eq.~\ref{eq:isca} with the Mie theory. 
We then compare the predicted scattered intensity 
(Eq.~\ref{eq:isca}), calculated using 
the classical MNR model and the WD01 model for $n(a)$, with our data.

\bfi

\begin{center}

\includegraphics[width=7cm,height=8.7cm,angle=90]{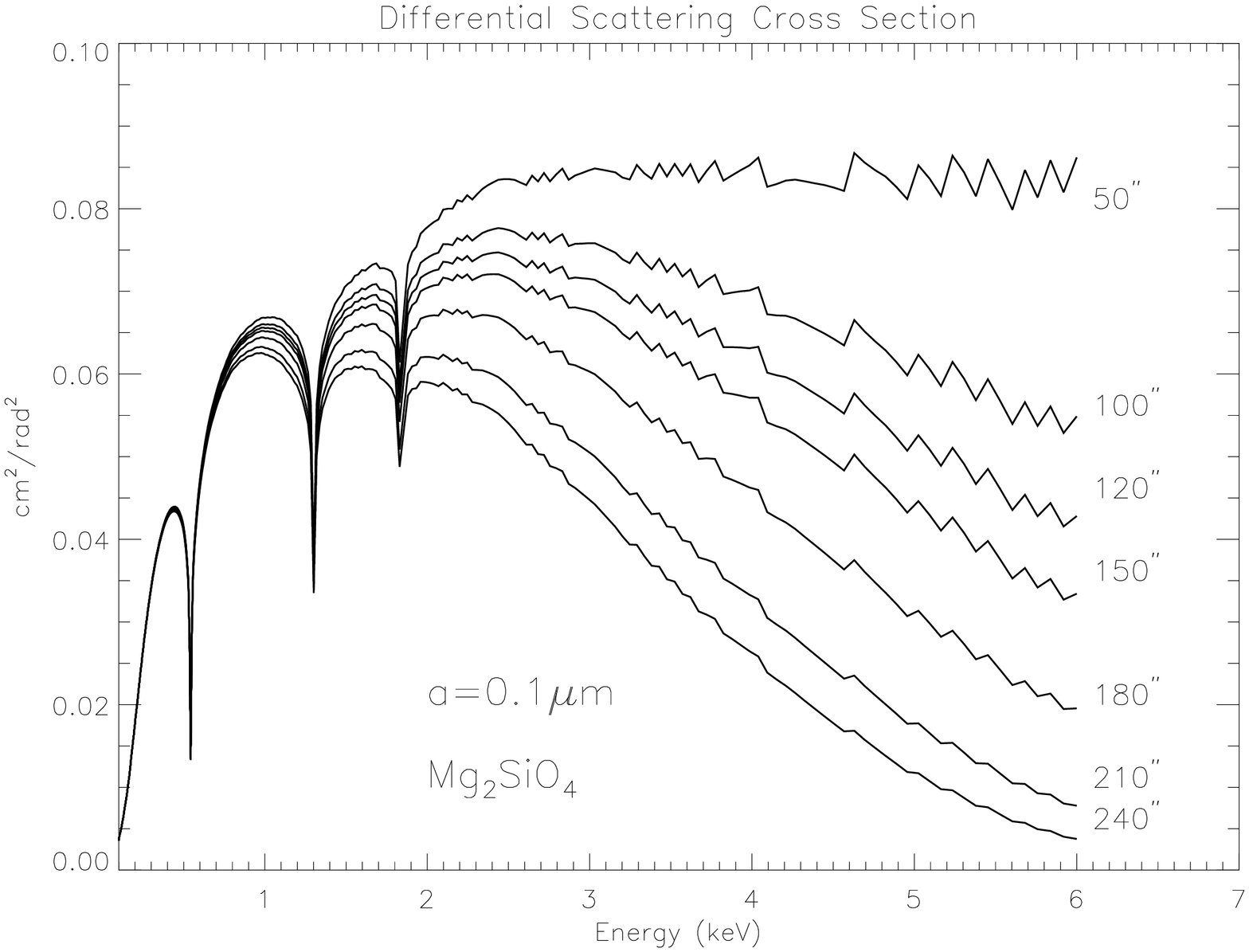}

\end{center}
\caption{\label{f:mie_i} The scattering cross section, calculated using the Mie theory, at specific scattering angles 
vs.\ energy for \mg. A single grain size of $0.1\,\mu$m is considered.}
\efi

\section{Data analysis}
Cyg~X-2 was observed by \xmm\ EPIC-pn 
\citep{2001A&A...365L..18S} in full-frame mode, 
EPIC-MOS \citep{2001A&A...365L..27T} cameras, and the RGS high resolution 
spectrometers \citep{2001A&A...365L...7D} on 
June 3rd 2002 for 
18.6 ks. {The central CCDs of MOS1 and MOS2 were
operated in timing mode and thus contain no imaging information}. 
We therefore show data of EPIC-pn and RGS only.\\    
We processed the data using the XMM-Newton Science Analysis Software 
(XMMSAS-5.4.1) as well as {\it ad hoc} routines not included in the standard software.

\subsection{\cy\ EPIC-pn spectral analysis}\label{pn_sou}

A single photon can lead to charges in a single
pixel (referred to as {\em single events}), two neighboring
pixels ({\em double events}), or three or four pixels (if the photon
hits a region close to the corner of a pixel). Pile-up can occur when two photons hit the same pixel in the same read-out cycle
(energy pile-up) or hit a neighboring pixel (pattern pile-up), leading to distortions in the pattern distribution of single and double events. 
The central pixels of the source image of \cy\ are heavily piled up. 
A diagnostic tool (XMMSAS task {\tt epatplot}) 
applicable to \xmm\ EPIC CCDs makes use of 
the event pattern distributions, which can be precisely 
modeled as a function of energy for X-rays passing through
the telescope in the absence of pile-up. 
From a comparison of such a modeled distribution with a {\it real} pattern distribution, 
the percentage of pile-up in a spectrum extracted from a given region can be estimated.\\  
Out-of-Time (OOT) events occur
during the read-out of a pn-CCD along the read-out direction. As
these events are accumulated only within a short time and are
distributed over the whole CCD length pile-up effects are a factor of 1000 smaller. 
After checking with the {\tt epatplot} procedure, we used these events 
to model the broad band spectrum (0.4-10 keV) of the central source. 
In principle the absolute flux can be evaluated from the OOT events. 
In the case of extreme pile up 
this is no longer possible, due to the ``pseudo-MIP" effect:  
If the charge within one pn-CCD pixel exceeds a threshold of
about 15 keV this event is regarded as being due to a minimum
ionizing particle (MIP) and all events in this CCD column and the
neighboring columns for this read-out frame are rejected on board \citep[pseudo-MIPs, ][]{2003CAL...WA..16F}.
However, in the case of very strong pile-up this threshold can be
triggered by normal X-rays. In extreme cases, in almost all
frames the columns at the center of the PSF are rejected. 
These spatial exposure variations are not fully reflected in the
event data files and in the XMMSAS software.
The ``pseudo-MIPs" have no appreciable influence on the source spectral shape; however, these
rejected columns affect the determination of the flux measured from the OOT events. 
Indeed, the pseudo-MIP rejection occurs preferentially in the
columns corresponding to the PSF core and therefore in the same columns
as the bulk of the OOT events, which then get rejected.
We modeled the spectrum extracted from the OOT events. Note that the source position, rather than the recorded position on the CCD, was used for the Charge Transfer
Inefficiency (CTI) correction using XMMSAS. The soft EPIC-pn spectrum is well fitted with a multi-temperature black body for the accretion
disk emission \citep{1984PASJ...36..741M} with $kT\sim$0.36\,keV at the inner radius,  
plus 
a comptonized black body spectrum for the emission of the neutron star \citep{1994ApJ...434..570T}. 
The soft (seed) photons have a 
temperature $kT_0\sim 0.8$\,keV  
before being Compton scattered 
to reach a temperature of $kT\sim 6.2$\,keV in an electron cloud of thickness $\tau$ (Tab.~\ref{t:epic_fit}). We found evidence of an emission line at energy 
$E\sim 6.7$\,keV, consistent with fluorescent emission by ionized
iron \citep[e.g., ][]{2002A&A...386..535D}. 
The soft spectrum is both absorbed by gas and dust and scattered by ID, i.e.\ light is deviated from our line of
sight ``subtracting" photons from the central source spectrum (PS95),  
We implemented in XSPEC a model for the scattering correction which is  based on handy empirical relations: 
$\tau_{\rm sca}=0.05\times N_{\rm H}-0.083$  and $A_V=0.56\ N_{\rm H} +0.23$, 
where $\tau_{\rm sca}$ is the scattering optical depth and 
\nh\ is in units 
of $10^{21}\ {\rm cm}^{-2}$. This relation is based on the study of 25 ROSAT sources (PS95) and it is not critically 
model dependent as long as we are dealing with a relatively small
correction for $\tau$.
The optical extinction $A_V$ value is only 1.3 for \cy\ \citep{1983ARA&A..21...13B}, 
therefore  the influence 
of scattering in the spectrum is practically negligible in the fit for such a low intervening column density (PS95). 

\begin{table}
\begin{center}
\caption{\label{t:epic_fit}Best fit parameters for the EPIC-pn spectrum of \cy\ in the energy band 0.4-10 keV, with a disk black body at temperature $kT_{db}$ 
\citep[{\tt DISKBB} in XSPEC, ][]{1984PASJ...36..741M} and a comptonized spectrum \citep[{\tt COMPTT} in XSPEC, ][]{1994ApJ...434..570T}, 
 affected by extinction of gas and dust \citep[{\tt TBABS} in XSPEC, ][]{wam00}. See text for the definition of the parameters. Errors are given
at 90\% confidence level for one interesting parameter.}
\begin{tabular}{lc}
\hline
\hline
$N_{\rm H}\ (\times10^{22}\ {\rm cm^{-2})}$		&		$0.19\pm 0.05$\\	 	
$kT_{db}\ {\rm (keV)}$				&		$0.36\pm 0.05$	\\ 		
$kT_0\ {\rm (keV)}$						&		$0.81\pm 0.03$	\\ 		
$kT\ {\rm (keV)}$							&		$6.2\pm0.1$	\\		 		
$\tau$									&	$1.62\pm 0.07$\\								
$E_{\rm Fe} {\rm (keV)}$								& $6.67 \pm 0.16$ \\
$\sigma_{\rm Fe} {\rm (keV)}$						& $0.2^{+0.3}_{-0.1}$ \\
$ EW_{\rm Fe} {\rm (eV)}$							& $73\pm 53$  \\
$\chi^2/dof$	& 1115/1111	 \\
\hline
\hline
\end{tabular}
\end{center}
\end{table}

\subsection{\cy\ RGS spectral analysis}\label{par:cy_rgs}
\begin{figure}[h]
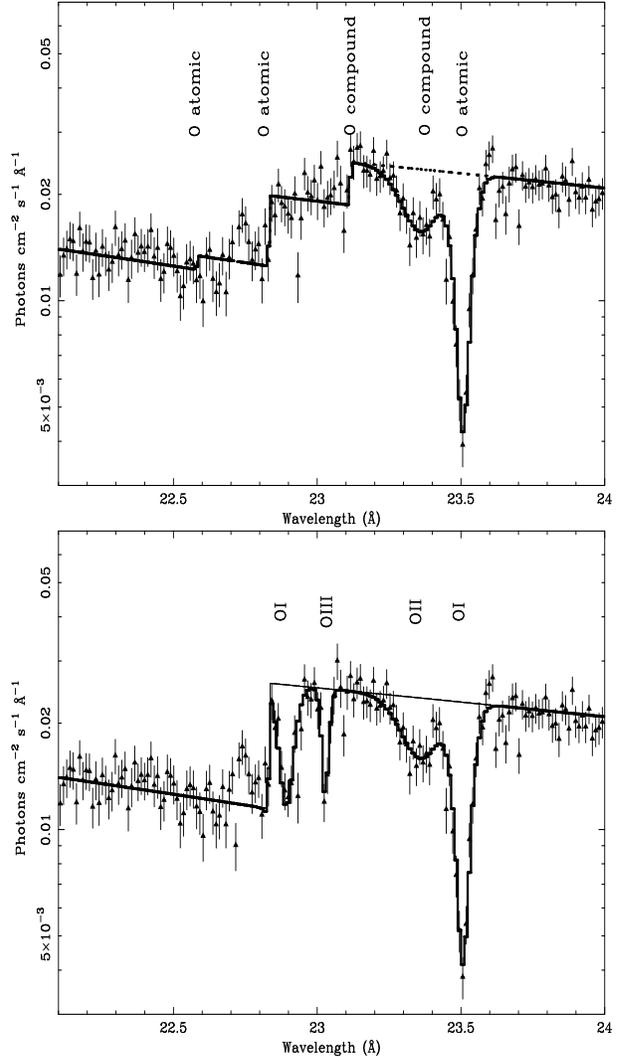

\begin{center}
{\includegraphics[width=7cm,height=8cm,angle=-90]{f2a.ps}}

{\includegraphics[width=7cm,height=8cm,angle=-90]{f2b.ps}}
\end{center}
\caption{\label{f:cy_oxi}Comparison 
between model~1 (top), and model~2 (bottom) used to fit the oxygen region, using the RGS data. 
Model~1 includes three absorption edges and two
absorption lines. Model~2 includes one absorption edge and four absorption lines.}

\end{figure}

As the RGS covers only the 0.35-2 keV band, the hard component cannot be constrained, therefore the 
thermal comptonization
parameters are fixed to the EPIC-pn best-fit values. 
The background spectrum contribution was evaluated using the ``blank field" observations specific for RGS. This guarantees the omission of any halo contamination in the
background. In the case of \cy, the halo contribution to the source spectrum is almost negligible, since the flux of the source is more 
than 20 times larger than the flux of the diffuse halo. 
We extracted the first and second order from RGS1 and RGS2 for a total of four data sets. The data
below 7\AA\ are affected by low and poorly calibrated sensitivity and are rejected.\\ 
Absorption by oxygen in the ISM could be studied in detail in the spectral region around 0.54 keV. 
Takei et al.\ (2003) 
found a complex structure for the oxygen edge region in a {\it Chandra}-LETG observation of \cy. In their analysis
they interpreted the spectrum in terms of absorption lines and edges from oxygen in both atomic and molecular form.
Although the RGS energy resolution is approximately 44\% less than LETG, we find similar complexity in the oxygen region: 
in particular a single oxygen edge at energy 
0.543\,keV is an unsatisfactory fit to the data. 
We first considered the approach of \citet{tak03} ({\em model~1} in Tab.~\ref{t:rgs_bothfit}). 
We included an additional edge 
in the 
fit, which improves the fit by 
$\Delta\chi^2/\Delta\nu=39/1$ (corresponding to a significance higher than 99.5\%). 
The two edge energies are fixed: 0.536 (23.13) and 0.543 (22.83) keV (\AA ), corresponding to compound and atomic oxygen, respectively. 
On the other hand, a third edge, strongly required by LETG data at 0.549 keV (22.58\AA, atomic oxygen), 
improves our fit only by $\Delta\chi^2/\Delta\nu=5/1$ (significance 97.5\%). 
At an energy of $0.524\pm0.003$ keV, consistent with the atomic oxygen $1s-2p$ transition, an absorption line of equivalent width (EW) $1.45^{+0.22}_{-0.14}$ eV 
is clearly detected. 
Finally, at $0.530\pm0.003$ keV some absorption line-like residuals, in addition to the known instrumental absorption line 
\citep{devries03}, still remain. 
Including at this position a second absorption line in the fit yields an energy that is interpreted as the $1s-2p$ 
transition of compound oxygen (Fig.~\ref{f:cy_oxi}, upper panel). 
The measured EW is $1.41\pm0.26$ eV, consistent with the findings of \citet{tak03}.\\ 
However, large uncertainties still remain in the laboratory measurements of oxygen bound 
with other elements and
the identification of such features is not conclusive. 
Other laboratory measurements 
\citep[e.g.,][]{mclk98,gmcl00}
of atomic oxygen around the K edge region would interpret the absorption structures 
as absorption lines from neutral and ionized oxygen. This is called {\em model~2} in Tab.~\ref{f:cy_oxi}. 
 The only two features in common with the 
Takei et al.\ 
interpretation are:
the absorption edge at 0.543 keV (22.83\AA ) and the $1s-2p$ transition line at 0.524$\pm0.003$ keV. The region between these two ``standard" features is fit by an absorption line at 22.89\AA , 
consistent with the 1s-3p transition of neutral atomic oxygen, and with another absorption line which would be consistent with a blend of unresolved lines of 
O{\sc iii}
(at 23.05 \AA ). Finally, the evident absorption line at 23.35\AA , also found by Takei et al., is interpreted as ionized atomic
oxygen 
(O{\sc ii}), 
as predicted by \citet{gmcl00} measurements (Fig.~\ref{f:cy_oxi}, lower panel). 
This interpretation ({\em model~2}) was also applied to \ch-HETG data of a sample of bright galactic 
sources \citep{juett03}. 
Such 
an 
ionized component would be interpreted as ionization of the ISM, localized in the vicinity of the source. In Tab.~\ref{t:rgs_bothfit}
the results of the two models are shown; there is no significant difference in terms of goodness of fit.

\begin{table}[h]
{\footnotesize
\caption{\label{t:rgs_bothfit}RGS fitting results for the oxygen region in Cyg~X-2.
 The parameters for the continuum emission are taken from the broad-band spectrum. The energies are measured in keV and
the Equivalent Width (EW) in eV.  
Model~1 fits the O vicinity with 3 edges
(two of which are from atomic O) and 2 absorption lines (one from atomic neutral O and the other from molecular O). 
Model~2 interprets the spectrum in terms of absorption by atomic O, either neutral or mildly ionized. Errors are given at 90\% confidence for one interesting parameter.}
\begin{center}
\begin{tabular}{lcc}
\hline
\hline
& model~1 & model~2  \\
\hline
{\bf Lines} & 	&\\
$E_1$ 	&   0.524$\pm$0.003      &   0.524$\pm$0.003		\\
$EW_1$		&   1.45$^{+0.22}_{-0.14}$        &   1.49$\pm$0.17		\\
$E_2$ 	&   0.530$\pm$0.001      &   0.530$\pm$0.001		\\
$EW_2$ 		&   1.41$\pm$0.26        &   1.44$\pm$0.28		\\
$E_3$ 	&   $\cdots$	         &   0.538$\pm$0.001		\\
$EW_3$		&   $\cdots$	         &   0.44$^{+0.19}_{-0.24}$	\\
$E_4$ 	&   $\cdots$	         &   0.541$\pm$0.001		\\
$EW_4$ 		&   $\cdots$	         &   1.01$^{+0.15}_{-0.26}$	\\

{\bf Edges} & & \\

$E_1$	&   0.536 fix.      	       &      $\cdots$  	\\
$\tau_1$	&   0.27$\pm$0.06   	       &      $\cdots$  	\\
$E_2$	&   0.543 fix.      	       &      0.543$\pm$0.001	\\
$\tau_2$	&   0.45$\pm$0.02   	       &      0.81$\pm$0.02	\\
$E_3$	&   0.549 fix.      	       &      $\cdots$  	\\
$\tau_3$	&   0.09$^{+0.07}_{-0.08}$     &      $\cdots$  	\\

\hline
\hline
\end{tabular}

\end{center}
}

\end{table}


\begin{figure}[h]
\begin{center}
\psfig{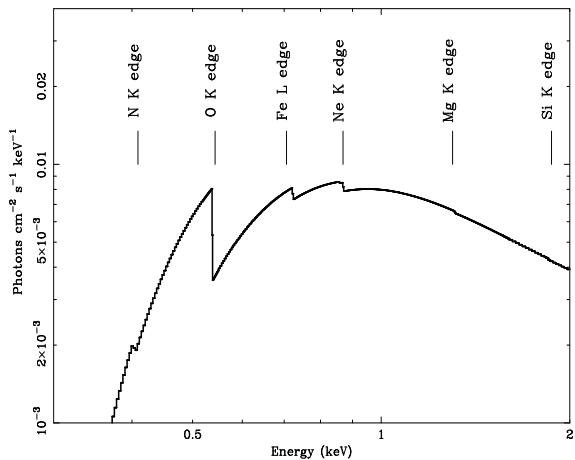}
\end{center}         
\caption{\label{f:abs}ISM absorption edges 
predicted for \nh 
$\sim2.2\times 10^{21} {\rm cm}^{-2}$ using the best EPIC-pn best fit model, evaluated from the OOT events data. Mg (1.3\,keV) and Si (1.84\,keV) 
are only marginally evident in the model.}

\end{figure}

The total equivalent hydrogen column density \nh $=(2.20\pm 0.02)\times 10^{21} {\rm cm}^{-2}$, as measured by the RGS, 
predicts K-shell absorption edges of nitrogen, 
oxygen, neon, and also iron L-shell, as
shown in 
Fig.~\ref{f:abs}. 
The statistics of the present data allow the 
determination of the physical parameters of the absorption. 
For each element $j$, we calculated the column density $N_j=\tau_j/\sigma_j$  using the photoelectric cross section 
computed 
from tabulated values 
\citep{1993ADNDT..54..181H,1995A&AS..109..125V}. The optical depth $\tau_j$ was measured from an absorption edge model. 
The edge model was applied to our best fit continuum, but fixing to zero the abundance of the element $j$ in the absorption model. 
Using the ISM abundances listed by \citet{wam00}, we derived the equivalent hydrogen column density.
In general we find an agreement within the errors between the column densities so derived and 
the \nh\ found in our best-fit model (Tab.~\ref{t:rgs_a}). 
We note that iron shows an overabundance of
$\sim20$\% compared to the \citet{wam00} ISM value.\\ 

\begin{table}[ht]
\renewcommand{\tabcolsep}{0.4mm}
\caption{\label{t:rgs_a}Relevant absorption edges in the RGS spectrum of \cy. 
The energy, the corresponding wavelength, and the optical depth ($\tau$) were obtained 
from the data. The
equivalent total hydrogen column 
density \nh\ was derived from the ISM abundances of Wilms, Allen \& McCray (2000). These are to be compared with the best-fit 
\nh $=(2.20\pm 0.02)\times 10^{21}\ {\rm cm}^{-2}$, measured by RGS. Errors are given at 90\% confidence level.}
\begin{tabular}{lcccc}
\hline
\hline
\multicolumn{1}{c}{Element} &
\multicolumn{1}{c}{Energy} &
\multicolumn{1}{c}{Wavelength} &
\multicolumn{1}{c}{$\tau$} &
\multicolumn{1}{c}{ $N_{\rm H}$} \\
& keV & \AA\ & & $10^{21}\ {\rm cm}^{-2}$\\
\hline
N K-shell & 0.409 (fixed)	& 30.31 & $0.15\pm0.06$ & $2.7\pm1.1$\\
O K-shell &   0.543 (fixed)  &  22.83  &  0.81$\pm0.10$ &  $2.8\pm0.8$\\
Fe L-shell &  $0.706\pm0.01$ & $17.56\pm0.02$ & $0.13\pm0.01$ & $2.8\pm0.2$\\             
Ne K-shell & $0.869^{+0.02}_{-0.06}$ & $14.26^{+0.10}_{-0.03}$ & $0.06\pm0.03$ & $1.8\pm0.9$ \\
\hline
\hline
\end{tabular}

\end{table}

\section{The analysis of the scattering halo}\label{halo_extr}

Thanks to the large effective area of the \xmm\ telescopes, coupled with the 
large field of view and spectral resolution 
of EPIC-pn, 
the ``pure" scattered radiation can be extracted from the halo spectrum. 
Thus  
for the first time we were able to analyze both the spatial and the spectral distribution of the scattering halo. 
The diffuse emission brightness is 
a 
few percent 
of the central source flux. The subtraction of
the 
\cy\ contribution relies on {\it (i)} an accurate understanding of the 
instrumental scattering by the surface of the gold mirrors of the telescope and {\it (ii)} a careful handling of the pile-up, 
affecting the central part of the source. 
Pile-up distorts both the spectral shape and the surface brightness profile (SBP).


\bfi[t]
\begin{center}
{\includegraphics[width=10cm,height=8.2cm,angle=0]{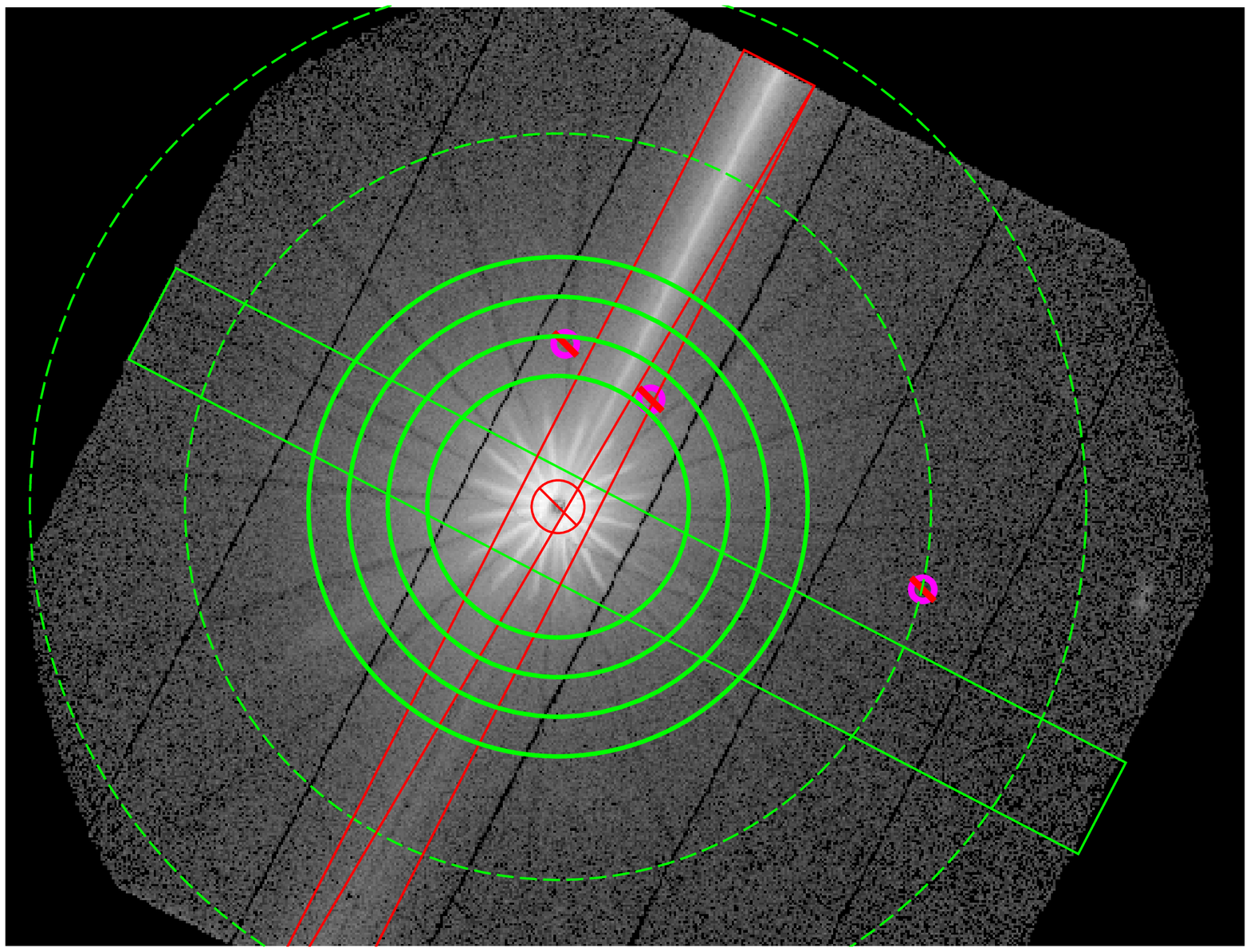}}
\caption{\label{f:cyg_image} EPIC-pn image of \cy\ in the 0.5-3 keV band. The solid line annuli show three of the extraction 
regions specific for the halo spectral analysis (Fig.~\ref{f:halo_sp_allrings}). The circles are drawn at 3.3, 4.3, 5.3, 6.3 arcmin. 
The dashed line annulus is the background. These regions were cut by 
the green rectangle across the detector (see \S~\ref{par:hspe}). For the halo spatial analysis (\S~\ref{par:hspa})
the annuli were instead logarithmically spaced. The barred regions were excluded from the data analysis: pile-up region (circle in the
center), OOT events (box), and the serendipity sources (points).} 
\end{center}
\efi

\begin{figure}[h]
\begin{center}
\psfig{file=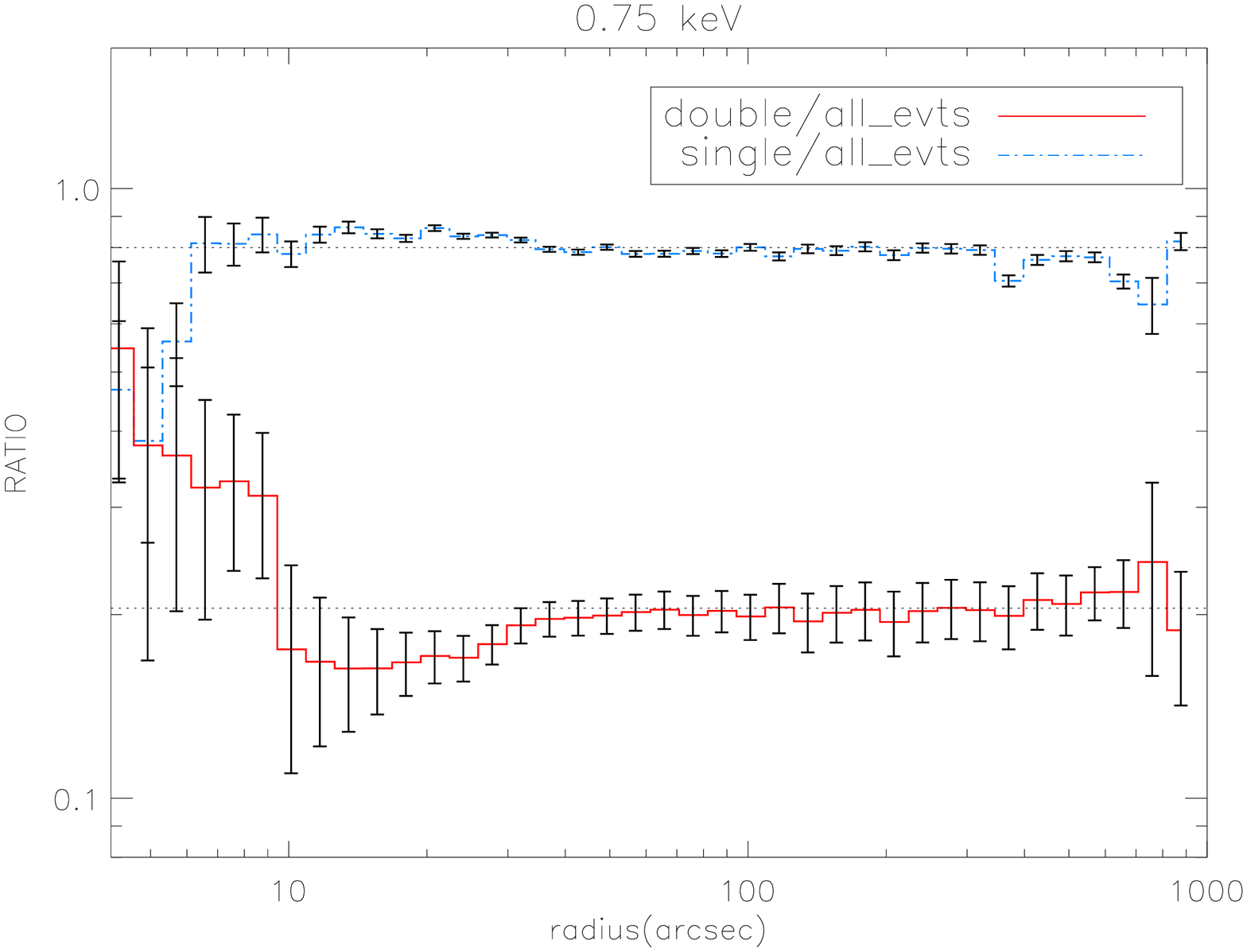,width=9.cm,height=7.cm,angle=90}
\end{center}         
\caption{\label{grade} Ratio between single events and all events (upper dash-dotted curve) 
and ratio between double and all events (lower solid curve), for EPIC-pn, around 0.75 keV. 
The dotted lines represent the values expected in absence of pile-up.}
\end{figure}

\subsection{Halo spatial analysis\label{par:hspa}}
As a preliminary step, all spurious sources were removed from the data set. 
The {\it Simbad}\,\footnote{http://simbad.u-strasbg.fr} search reports 60 objects in the field of view of \cy\ of which only 3 are known X-ray sources.  
These were not detected in our observation, but nevertheless the data at those positions were discarded (see Fig.~\ref{f:cyg_image}). 
The SBP of the scattering halo was evaluated extracting the photons 
at selected energies between 0.4 and 2\,keV in bins of 0.25 keV from logarithmically spaced annuli centered on the source, 
excluding the emission coming from the 
read-out streak due to OOT events. 
For the background we took an annulus from $9.5^{\prime}$ to $13^{\prime}$, avoiding the 
read-out  
streak region. The chosen background is consistent with the scaled value of the background taken from the ``blank field" data for EPIC-pn \citep{2003A&A...409..395R}.
For the extraction we 
took into account the different exposures of each quadrant of the detector selecting just the time intervals
 when all four quadrants were on at the same time. Any excluded region (serendipity sources, OOT events, regions outside the chip boundaries) was of course taken into
 account in the computation of the extraction area.\\ 
Pile-up is close to 100\% 
at the center of the source. This causes the characteristic ``hole" in the 
spatial profile, but to a lesser degree also distorts the profile shape up
to many arcsecs from the source. In order to evaluate the dependence of pile-up as a function 
of the distance from the source, we extracted the radial profile of the source at
different energies for single events, double events and the total events, selecting different {\it pattern} from the data. 
 Where the
count rate is low
(i.e.\ no pile-up) the ratio of the radial profiles extracted with these different {\it patterns} (single, double and all events), 
should be a constant value (Fig.~\ref{grade}). We see that for single events pile-up affects the profile up to $\sim$40$^{\prime\prime}$; thus we 
studied the data only outside this radius.
We divided the photon histogram by the exposure map, and the areas of the annuli, that was 
also corrected for the zones excluded in the photon extraction. Moreover, each photon is vignetting corrected by the ratio of
the effective area at the aim point of the telescope and at the position where the photon itself is detected.   
The resulting radial intensity distribution is now in units of $cts/s/arcsec^2$.
The next step is the subtraction of the PSF from the data. 
The model for the \xmm\ PSF as a function of energy and off-axis angle was derived by the analysis of 110 point-like sources \citep{ghi02}. 
The instrumental PSF is described by a King profile:
\begin{equation}
PSF\propto\left(\frac{1}{[1+(r/r_c)^2]^{\alpha}}\right),
\end{equation}
where $r_c$ is the radius of the PSF core, and $\alpha$ the slope of the profile. 
The energy dependence\footnote{The off-axis dependence of the PSF is neglected as the sources we study are located on-axis.} 
for these parameters is $\alpha=1.525-0.015\times E$ and $r_c=6.636-0.305\times E$. For a full description of the PSF modeling see \citet{ghi02}.  
For each energy we considered in the halo spatial analysis, 
we used the appropriate King profile.  
We compared this PSF profile with the calibration target \object{Mrk~421} (Fig.~\ref{f:prof_king}), which 
will be extensively used in this paper as a halo-free source, in comparison with \cy. 
This bright point-like source was observed in a special ``masked" mode. In this submode the
central $11 \times 11$ pixels
(of size 4.128 arcsec each, i.e.\ a square of $\sim 45$ arcsec
or a half diagonal of $\sim 0.53$ arcmin)
were set as bad pixels and thus masked on board.
This prevents the creation of
pseudo-MIPs (\S~\ref{pn_sou}) and thus the OOT events can be used to recover the
source flux value as well as the spectral shape.
Thus the Mrk~421 profile, also in units of $cts/s/arcsec^2$, could be easily normalized to the raw count rate ($cts/s$) of the source in a given energy bin.  
\bfi[t]
\begin{center}
{\includegraphics[width=7cm,height=8cm,angle=90]{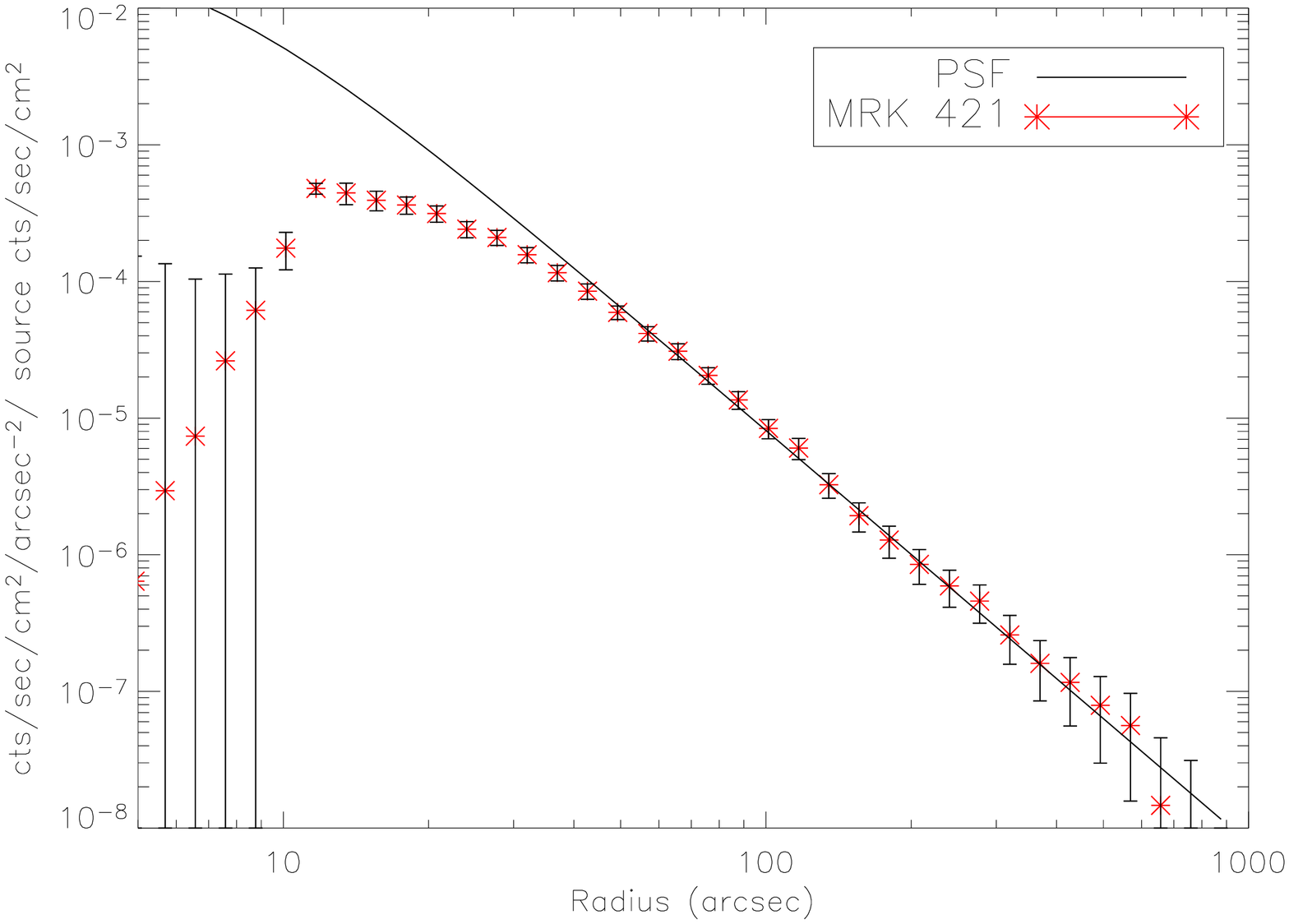}}
\caption{\label{f:prof_king}Comparison between the PSF-model (solid line), 
Mrk~421 (asterisks) between 0.9 and 1 keV. For radii smaller than $\sim$30$^{\prime\prime}$, the flatness of the profile is mainly due to the
mask used in the observation. The counts in this region are non zero due to a slight misalignment of the mask with respect to the source coordinates in this particular
observation.} 
\end{center}
\efi

Such a normalization was not straightforward for 
the \cy\ profile, since the central pixels of the source image 
are unusable due to 
the 
strong pile up. However, 
since the absolute level of the Mrk~421 surface brightness profile (in units of $arcsec^{-1}$) is known, 
we could safely normalize the \cy\ halo radial profile. We made the two profiles
overlap in the interval between $\sim$40$^{\prime\prime}$-100$^{\prime\prime}$, 
where \cy\ does not show
significant scattered extended emission and therefore the profile is
dominated by the instrumental PSF. We have seen above that the PSF profile slowly changes with energy, hence 
overlapping the source profile with the PSF considering a too
large energy band would introduce additional uncertainty. For the largest energy bin we considered (0.25 keV, \S~\ref{par:mo_profile}) the 
net uncertainty on the slope $\alpha$
is $\pm$2\%. As the halo analysis is confined to radii $>40^{\prime\prime}$, the uncertainty introduced by $r_c(E)$ is negligible.   
We note that in principle, to evaluate the spectral 
flux we could use either the OOT events or the RGS high resolution spectrum. 
The uncertainties in the cross calibration between PN and RGS flux may reach 20\% and make the RGS data inappropriate for this purpose.   
The OOT events were also unusable for the normalization due to the pseudo-MIP effect (\S~\ref{pn_sou}).\\ 
At the end of the procedure just described, the resulting SBP of \cy\ , which is the summed contribution of an extended emission 
and the instrumental profile, could be subtracted by the PSF and then modeled
(\S~\ref{par:mo_profile}).  

\subsection{Halo spectral analysis}\label{par:hspe}
\begin{figure}
\begin{center}
{\includegraphics[width=6.7cm,height=9.2cm,angle=90]{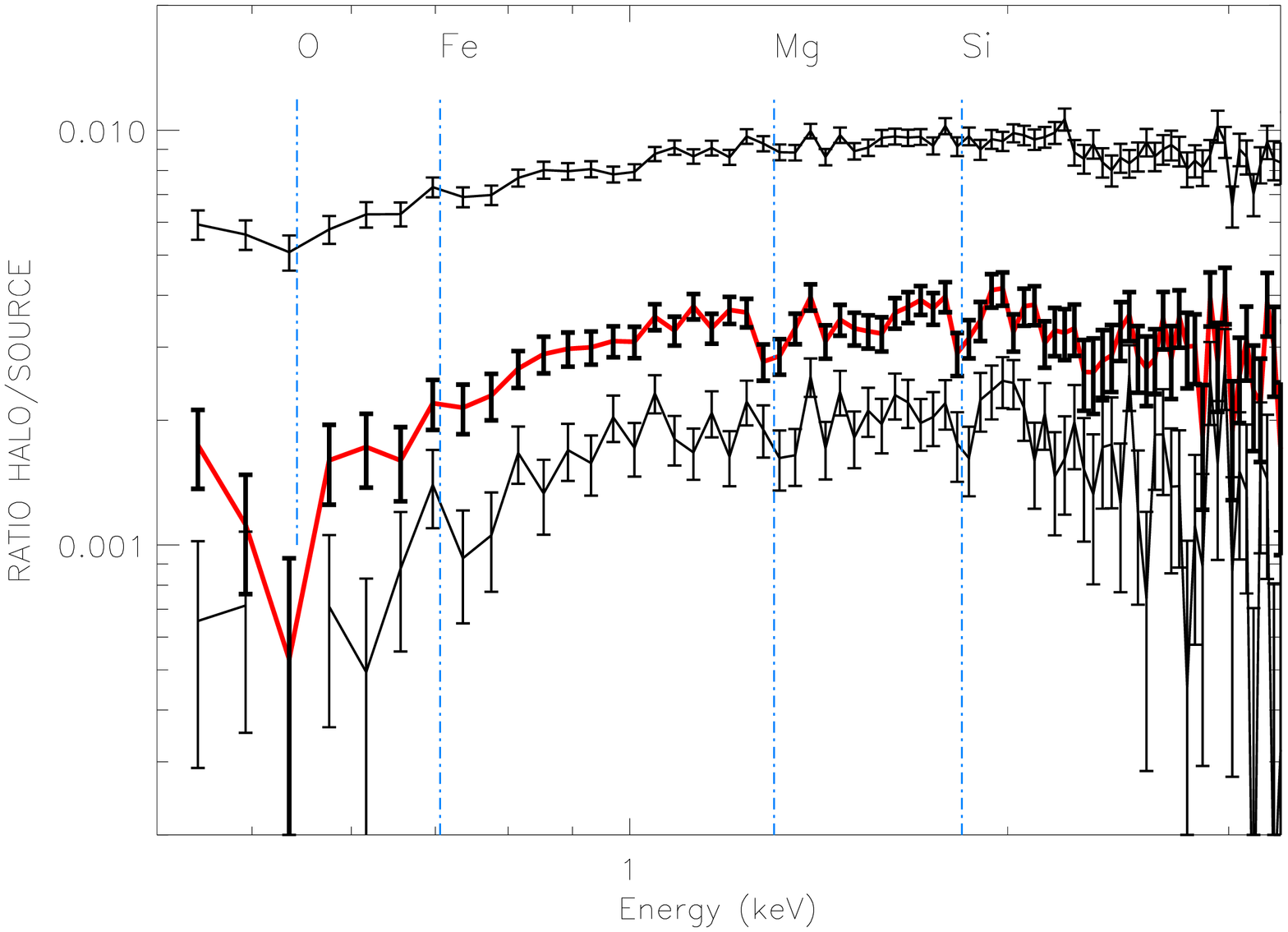}}
\end{center}         
\caption{\label{f:halo_sp_allrings}
The ratio between the halo spectra and the central source spectrum. At this stage the data are still convolved with the PSF. The spectra were extracted from annuli  
--from top to bottom-- centered at 3.8, 4.8, and 5.8 arcmin respectively (Fig.~\ref{f:cyg_image}). 
At larger radii the PSF contribution becomes less important. Dashed vertical lines: energies of relevant ID elements (O, Fe, Mg, Si). Error bars are 68\% confidence level
(1$\sigma$). The vertical axis units are arbitrary as the absolute flux of the source could not be recovered from the OOT events (\S~\ref{pn_sou}).}
\end{figure}

For the study of the spatial variation of the spectrum
we first selected radial distances from the source from $1.3^{\prime}$ to $7.3^{\prime}$ divided in annuli of  $1^{\prime}$.  
The background was extracted from the same region as for the halo spatial analysis.
A spatial selection\footnote{the 
RAW CCD coordinate selection was: RAWY=$180-199$. This RAWY range always contains the source 
if it is observed on-axis and is well calibrated, and allows us to avoid the uncertainties related to the RAWY dependence of the energy response.}, 
was applied to ensure that the energy response   
stays constant across the detector.
In order to have a model independent estimate of how the halo spectra may change as a function of the distance from the source, 
we normalized the spectra in the rings (in units of raw detector counts) to the source spectrum 
estimated from the OOT events (\S~\ref{pn_sou}). In this way, first, all the features belonging 
to the X-ray emitter itself cancel out 
and, second, assuming 
that absorption by ID is rather uniform within the few arcmin across the halo, also the absorption component 
is eliminated from the halo spectrum. The result of this procedure is shown in Fig.~\ref{f:halo_sp_allrings}. 
The spectra of the \cy\ halo 
are plotted in order to illustrate the behavior of the halo with increasing angular distance from the source, 
as the PSF contribution becomes less important. The extraction regions of these spectra are marked in Fig.~\ref{f:cyg_image}. Here the inner radii range from 3.3\pr\ to 5.3\pr, where the halo is more relevant.
Note that at this stage, the instrumental mirror scattering spectral energy distribution is still to be subtracted. The vertical axis of Fig.~\ref{f:halo_sp_allrings} 
is in arbitrary units as the absolute normalization of the source spectrum cannot be recovered from the OOT events (see \S~\ref{pn_sou}). 
Below 2 keV, a decline of the curve is evident and the presence of three absorption features corresponding to the energy of O 
(0.54 keV), Mg (1.3 keV) and Si (1.84 keV) are observed. 
The vertical dashed lines indicate the position of these features and for completeness, the Fe~L edge position (0.709 keV) is also marked. 
In the absence of any extended emission (i.e.\ for a point-like source) 
the result of the above exercise on the spectrum extraction is the pure instrumental mirror scattering. This was done for the point source 
Mrk~421, to which exactly the same procedure was
applied (Fig.~\ref{f:mkn_scatt} for an example at radius 180\pr\pr). 
Since the point-source data are real data (though with high statistics), they are still affected by noise. 
Thus, to avoid loss of information on the halo spectra, the Mrk~421 spectrum was smoothed before 
subtracting it from the data. 
The PSF spectrum does not change significantly as a function of the scattering angle at soft energies, keeping its almost constant shape between 0.4$-$1.2\,keV. 
As we do not have precise information on the absolute flux of \cy, in order to perform
the PSF subtraction we made the halo spectrum collected at each annulus coincide with the homologous spectrum of Mrk~421, 
in the spectral region where the emission is
due to scattering by the telescope mirrors only ($>3.5$ keV). The PSF-subtracted scattered radiation is shown in Fig.~\ref{f:data_mg2sio4} for an observed angle of 4.8\pr. 
From the lowest observed energy up to the
silicon edge energy, the PSF subtraction is not a concern, as the extended emission is well above the PSF spectrum for a wide range of scattering angles 
(see Fig.~\ref{f:cy_prof} for an example at 1 keV). The comparison
PSF-halo is less straightforward for the Si feature and the continuum just below 1.84 keV, as the scattered emission becomes weaker with respect to the PSF. 
In this region, a small difference in the chosen normalization may affect the results of the halo modeling. 
The PSF subtraction has the effect of making the Si feature appear even deeper than what
already observed from the raw data (Fig.~\ref{f:halo_sp_allrings}). If, for example, the normalization of the halo spectrum were $\pm10$\% of what we calculated, 
the error bars for the Si feature in the PSF-subtracted spectrum (Fig.~\ref{f:data_mg2sio4}) would increase of an additional
6\%, considering that the ratio between the PSF and the raw data at this position is $\sim30$\% at $\theta_{\rm obs}=4.8^{\prime}$.

\bfi
\begin{center}
{\includegraphics[width=6.5cm,height=8cm,angle=90]{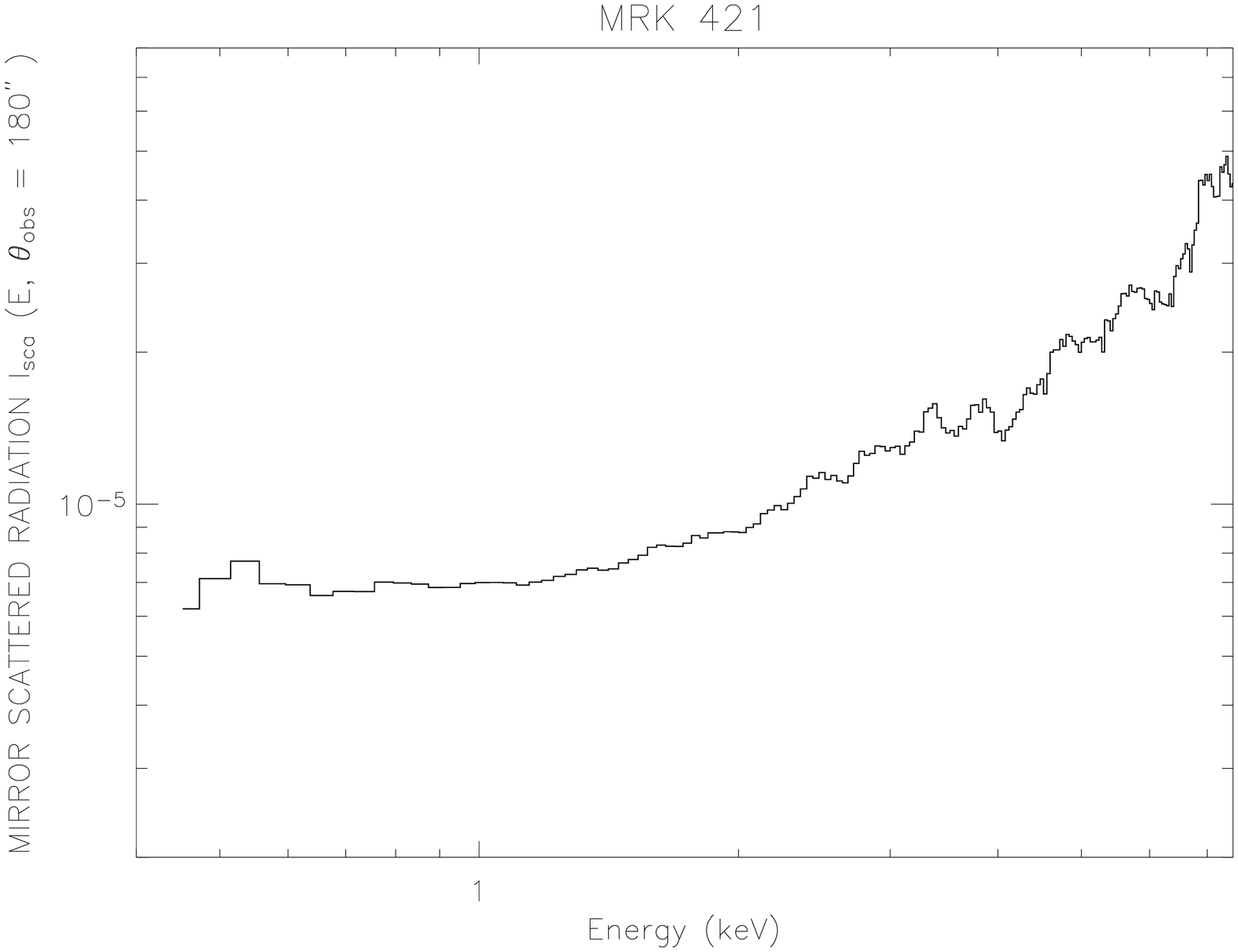}}
\caption{\label{f:mkn_scatt}XMM EPIC-pn pure-scattering mirror spectrum at a radius 180$^{\prime\prime}$ from the source, extracted from the halo-free source Mrk~421.
 We note a smooth increase of the scattering above 2 keV for
larger radii.} 
\end{center}
\efi


\section{The halo modeling}\label{par:hmo}
There are two critical terms in 
Eq.~\ref{eq:isca}. One is the differential scattering cross section and the other is the assumed grain size distribution. 
For both the spatial
and the spectral modeling of the scattering halo, we used the differential scattering cross section calculated using the Mie theory. 
For each value of
the parameter $X=2\pi\ a /\lambda$, where $a$ is the grain size and $\lambda$ is the wavelength of 
the incoming wave, we calculated the parameter  $d\sigma/d\Omega$ for 115
 scattering angles $\theta_{\rm sca}$ from 0 to 3$^{\circ}$. The scattering cross section depends also 
on the diffraction index $m$ which is determined by the material and is energy dependent. 
In this study we use the tabulated values of $m$ for graphite, olivine, and pyroxene 
as determined by \citet{1993ADNDT..54..181H} except in one case (\S~\ref{par:chem}), 
when we test the specific ID composition proposed in D03. In that case, the calculation for $m$ takes
into account the XAFS near the edges.\\   
The code\footnote{ftp://climate.gsfc.nasa.gov/pub/wiscombe/Single\_Scatt/ \\ Homogen\_Sphere/Exact\_Mie/} \citep{1980ApOpt..19.1505W} 
used to generate the scattering cross sections is reliable for $X<2\times10^4$. For large values of $a$ 
(say $a>1.2 \mu$m for $E=2$\,keV), the anomalous 
diffraction theory should be used \citep{1957lssp.book.....V}. We ignore very large grains in our calculation since, as shown in 
\citet{drainetan}, 
radii $>0.4 \mu$m contribute less 
than 1\% at large scattering angles, and less than 20\% at $\theta_{\rm sca}<100^{\prime\prime}$. In bright sources, like \cy, 
pile-up hampers the possibility to investigate the halo at small scattering angles, where
the effect of very large grains, or grains located very close to the source may be relevant \citep{pk96}. We also restricted our modeling to energies $<$2 keV, 
above which the halo
contribution drops dramatically in the case of \cy.\\       
Thus, the chosen grain size interval is 0.005-0.25 $\mu$m or 0.00035-0.8 $\mu$m 
when the MRN dust size distribution model or WD01 model is adopted, respectively.
The dust size intervals were divided in 200 logarithmically spaced size bins.  
We allowed the power law index of the MRN distribution to vary by 20\% around the typical value 3.5.  
WD01 tested their grain size
distribution for two different values of the ratio of the total over selective optical extinction: $R_V$=3.1 
and 5.3, and for different carbon abundances. \cy\ is located at
galactic latitude $b= -11.3^\circ$ 
where the ISM is diffuse \citep[no CO detected, ][]{2001ApJ...547..792D}, thus we considered $R_V$=3.1. 
We used the set of parameters for slopes and coefficients of the
dust distribution corresponding to a carbon abundance in PAH alone of 6$\times$10$^{-5}$ \citep[Tab.~1 of ][]{wd01}. 
For both models, the lower and upper limit of the integral on the dust distribution parameter $x$ were left as free parameters. 

\subsection{Spatial Modeling of the Halo}\label{par:mo_profile}

\begin{figure}[ht]
\begin{center}
\includegraphics[width=7cm,height=8cm,angle=90]{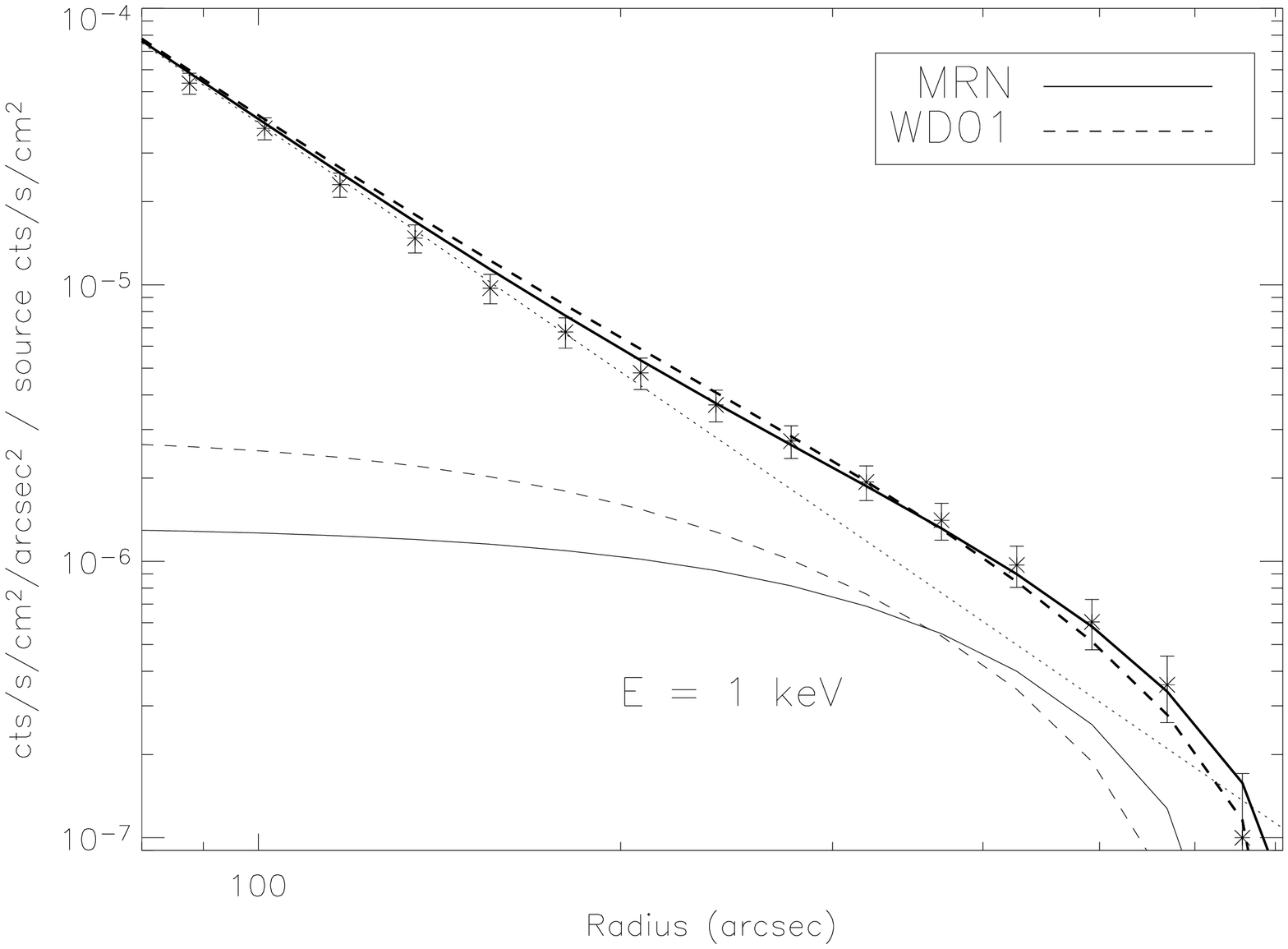}
\caption{\label{f:cy_prof}The data (halo+PSF) (asterisks) compared to the PSF (dotted line) around 1 keV. 
The dashed and the solid thin lines are two different halo models (MRN and W01), while the solid and dashed thick 
lines indicate the best-fit to the total data (model+PSF) relative to MRN and WD01, respectively.}

\end{center}
\end{figure}

The SBP was extracted and subtracted from the PSF contribution as described in \S~\ref{par:hspa}. At a fixed energy, the model has three free parameters 
(Eq.~\ref{eq:isca}): $d\sigma/d\Omega$, $x$, and $n(a)$.  
The best fit was reached through $\chi^2$ minimization. We considered rays only scattered once before being observed. Double scattering occurs for optical depths $\tau_{\rm sca}$ 
close to unity, indicating a very high dust column density \citep[PS95, ][]{cos04}, which is not observed toward \cy . 
In Fig.~\ref{f:cy_prof}, the \cy\ SBP of the halo at 1 keV is shown. 
We tested the MRN and the WD01 for the dust size distribution. Both
models provide an acceptable fit in terms of $\chi^2$ ($\chi^2_{red}$=1.29 and 1.33, respectively) .  
The WD01 distribution spans a wider range of grain sizes. In particular, scattering by grains with size $a$ in the range $0.25-0.4 \mu$m 
have the effect of enhancing the halo at smaller radii ($\ltsim 200^{\prime\prime}$).  
The intensity of the halo is parameterized by the scattering optical depth $\tau_{\rm sca}$, defined as: 
$I_{\rm frac}=I_{\rm halo}/I_{\rm tot}=1-e^{-\tau{_{\rm sca}}}$, where $I_{\rm halo}$ is the 
flux of the scattered emission, and $I_{\rm tot}$ is the total source emission (PS95). At 1 keV we measured 
$\tau_{\rm sca}$= 0.054$\pm$0.018 and 0.067$\pm$0.018 for the MRN or the WD01 model, respectively. The error quoted here is statistical and 
does not include any uncertainty in the
background subtraction. 
The best fit shows a minimum and maximum value for the fractional path at which the scattering occurs, $x$, 
of $\sim$0.001 and $\sim$0.6, respectively. In this interval the halo profile shows a smooth 
distribution of dust along the line of sight up to a fractional distance $x\sim0.6$, after which the halo is unaccessible due to pile-up. 
However, if the upper limit of $x$ is constrained to be close to 1 
(we fixed it at 0.99) indicating that we
are actually observing scattering occurring at all distances, the fit worsens 
for both the WD01 and MRN models ($\Delta\chi^2=37$ for $\Delta\nu=1$, corresponding to a significance $>99.5$\% ).
We then extended this analysis to the energy range at which the halo is observable. 
In Fig.~\ref{f:tot_tau} we show the total scattering optical depth, derived from the SBP using energy intervals of 0.25 keV, in the energy range
0.4-1.9 keV. The large bin size smoothes out any features, leaving just the general shape the spectral energy distribution of the scattering optical depth. 
The solid line in Fig.~\ref{f:tot_tau} refers to
the theoretical value of $\tau_{\rm sca}$ as predicted by D03 at the mean energy of the extraction bin. 
This was derived by multiplying the theoretical value of $\sigma_{\rm sca}$ (D03) with the hydrogen column density toward \cy\ that we measure.

\bfi[h]
\begin{center}
{\includegraphics[width=8.5cm,height=7cm,angle=0]{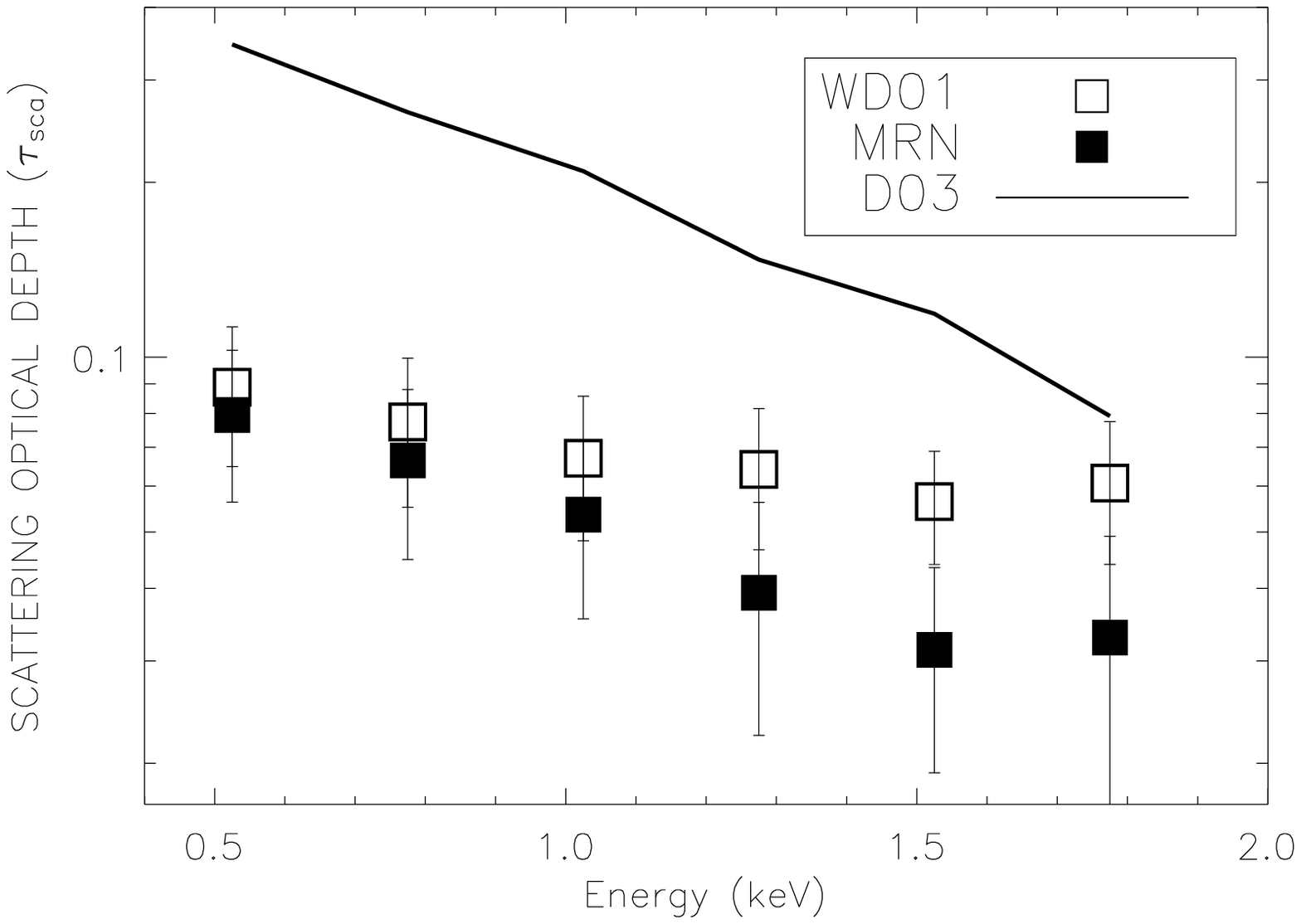}}
\caption{\label{f:tot_tau} The scattering optical depth as a function of energy, as measured from the SBP of \cy\ halo for the MRN model 
(filled squares) and WD01 (empty squares). The data were collected in energy bins 0.25 keV wide. 
The solid line refers to the value of the total $\tau$ at the center of the bin, as predicted in D03. 
  } 
\end{center}

\efi


\subsection{Spectral Modeling of the Halo}
\begin{center}
\begin{figure}[t]
\psfig{file=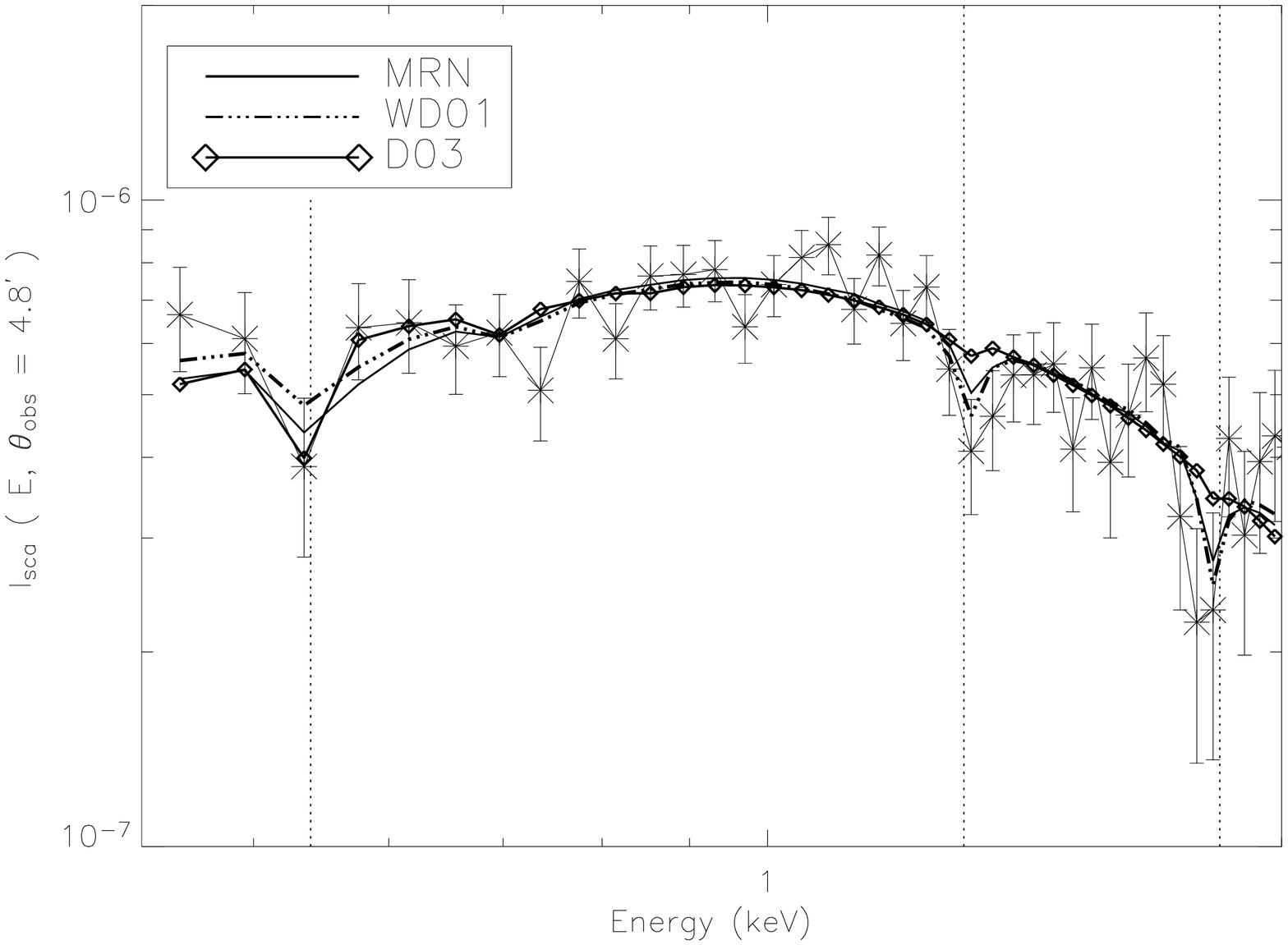,width=9.5cm,height=7.5cm,angle=90}
\caption{\label{f:data_mg2sio4}EPIC-pn data of the Cyg~X-2 halo, extracted at $\sim$4.3$^{\prime}$ from 
the source and PSF subtracted (asterisks), 
compared with the best-fit models: WD01 (dashed dotted line), MRN (solid line), and D03+WD01 model (diamonds).
O-K, Mg-K, and Si-K edge energies are indicated by dotted lines.}
\end{figure}
\end{center}

\begin{center}
\begin{figure}[t]
\psfig{file=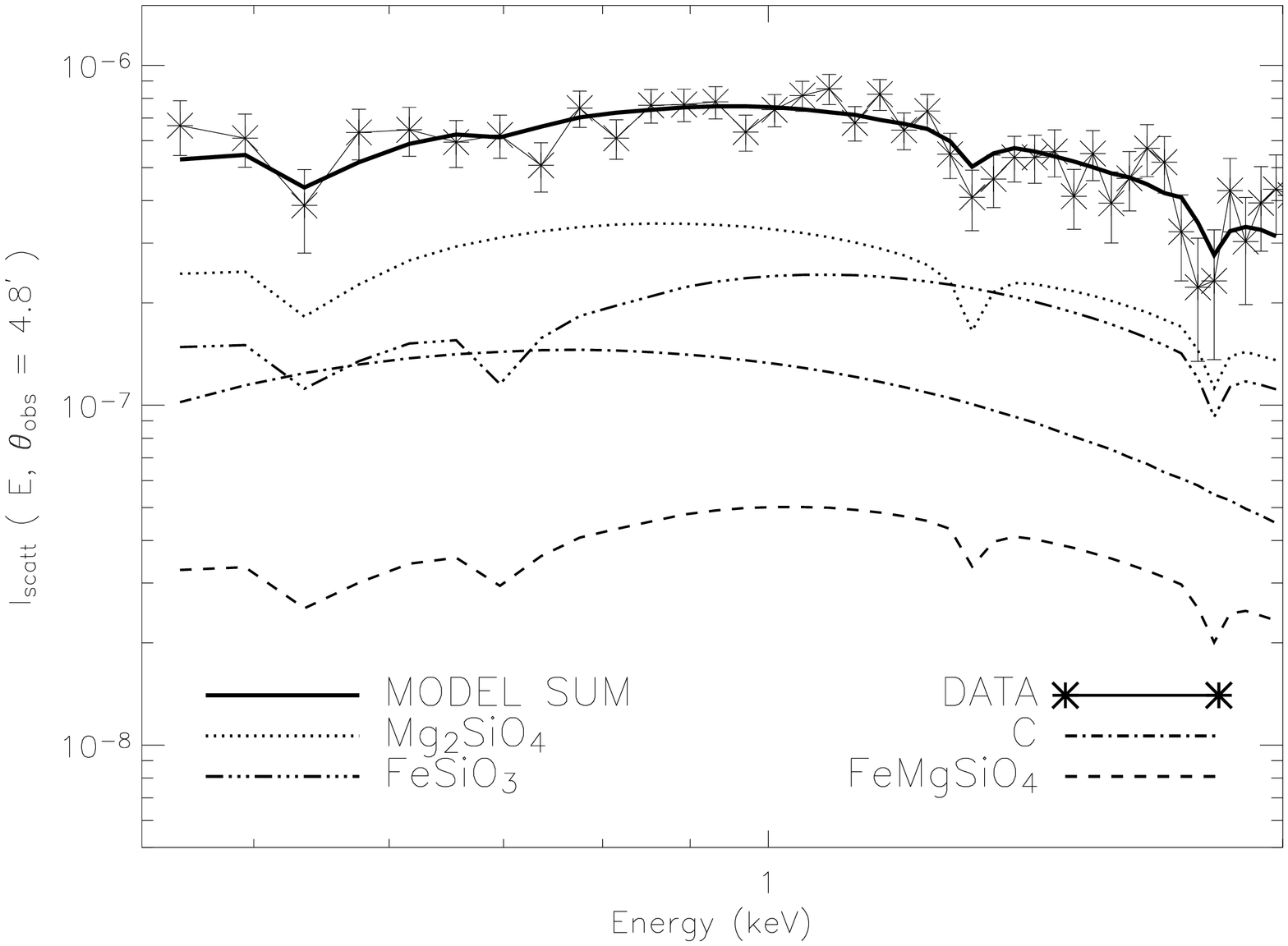,width=9.5cm,height=7.5cm,angle=90}
\caption{\label{f:mrn_contr}
Relative contributions to the best-fit model, using the MRN distribution. The data are the same as in Fig.~\ref{f:data_mg2sio4}.}  
\end{figure}
\end{center}

In Fig.~\ref{f:data_mg2sio4} we show the pure scattered radiation energy distribution (i.e. PSF subtracted), collected in a ring centered at 4.8$^{\prime}$. 
The subtraction of the mirror scattering and of the
central source contribution was described in \S~\ref{halo_extr}. We avoid the data
below 0.4 keV as the 
effective area calibration 
may still suffer from uncertainties in this range. 
At soft energies, the halo spectra collected in different annuli, shown in Fig.~\ref{f:halo_sp_allrings},    
do not vary dramatically (within statistical errors) among each other after the PSF subtraction. 
Above 2 keV the differential scattering cross section does change dramatically as a function 
of the scattering angle (Fig.~\ref{f:mie_i}); however the contribution of the PSF wings is dominant at this energy and 
this change is unobservable in the present data. 
In principle, after the PSF subtraction, the modeling of the halo spectra, at each angular distance, 
would reveal the local properties of the dust grains. Indeed, inhomogeneity in the dust spatial distribution, size of the dust particles, 
and chemical composition would result in a
spectral change. Unfortunately, these changes are tiny with respect to the instrumental uncertainties.
The modeling here refers just to the ring around $4.8^{\prime}$, where the halo is brighter and thus the statistics are highest.
The spectral features are still very well observable after the PSF subtraction. 
If artificially fitted with Gaussian profiles, the significance is 2.3$\sigma$, 3.2$\sigma$, and 4$\sigma$ for O, Mg, and Si,
respectively. 
Although the 
dust features of oxygen, magnesium and silicon are clearly detected in the scattered emission, still the statistics do not allow a unique interpretation of the data. 
In the attempt to give a quantitative picture of the observational evidence, we chose a model 
with
only the dominant well-known compounds of the diffuse interstellar dust environment. 
We calculated the scattered intensity 
expected at the chosen angular distance, for the most common 
constituents of interstellar grains, 
i.e. graphite (carbon) and silicates. 
Silicates are found mostly in the form of Mg$_{2x}$Fe$_{2(1-x)}$SiO$_4$ and MSiO$_3$, where M is either Mg or Fe,  
thus we included olivine in the three different forms of MgFeSiO$_4$ ($\rho$=3.8~g~cm$^{-3}$), 
Fe$_2$SiO$_4$, ($\rho$=4.39~g~cm$^{-3}$) and Mg$_2$SiO$_4$ ($\rho$=3.27~g~cm$^{-3}$). Pyroxene are 
in the form of MgSiO$_3$ ($\rho=$3.2~g~cm$^{-3}$) and FeSiO$_3$ ($\rho$=3.8~g~cm$^{-3}$). 
For a given dust size distribution model (MRN or WD01), the relative contributions of the
different grains materials are left as free parameters. 
Moreover, we considered also a model where only MgFeSiO$_4$ is taken as representative
compound for silicates (D03).   
For this model, 
the dust size is distributed according to WD01, and in the scattering cross sections XAFS are included.   
The goodness of the fit was evaluated 
by minimizing $\chi^2$.\\ 
The data are ignored below 0.4 keV, therefore we do not expect to put any constraint on the carbon component, whose main feature is at 0.28 keV. 
However, the carbon component is included in the fit since it may influence the shape of the continuum at energies above the carbon edge. Carbon contribution was allowed
to vary between 20\% 
 and 30\% of the total amount of dust \citep{whi03}, for all the models we tested.
The data are
modeled by a mixture of  silicate and pyroxene. In Fig.~\ref{f:data_mg2sio4} we display 
the three dust models tested. A mixture of silicate compounds (labeled MRN and WD01) seems to acceptably interpret the data.
C, O, Fe, Mg and Si alone account for 95\% of the dust components. All elements except O account for 15-30\% of 
the total amount 
of dust, in various forms \citep{whi03}. 
Here we assume that these elements describe 100\% of the observed scattering. 
The linear combination of the compounds contribution is shown in Fig.~\ref{f:mrn_contr} for the MRN distribution. In the WD01 case, the combination of compounds is not
significantly different from MRN (Fig.~\ref{f:data_mg2sio4}). The relative contribution to unity is Mg$_2$SiO$_4$=0.42, C=0.26, FeMgSiO$_4$=0.054, and FeSiO$_3$=0.25. 
Fe$_2$SiO$_4$ and MgSiO$_3$ contribute for a negligible fraction ($<$0.001). 
Counting
the contributions of the single elements, we obtain that roughly 42\% of O, 26\% of C, 11\% and 13\% of Si and Mg, and 5\% of Fe are needed to fit the data. 
\section{Discussion}


\subsection{The chemistry of dust grains}\label{par:chem}

For the first time the signature of the elements locked in dust grains and responsible for the scattering of the X-rays is detected. 
The best fit indicates a major contribution by olivine and pyroxene. 
We cannot exclude the presence of other different compounds. Being
 the best fit a linear combination of scattered intensities for a given compound, adding too many components would 
 not be necessarily a true physical interpretation of the data.
In principle, the depth of the spikes in the scattered spectrum tells us about the intrinsic properties of the dust grains. 
Magnesium and silicon are not detected but marginally in absorption (Fig.~\ref{f:abs}), while they are prominent features in the scattered spectrum, 
even in the raw data at the angles where 
the statistics is
maximal (Fig.~\ref{f:halo_sp_allrings}).  
Only a deeper observation will allow us to quantify this possible
discrepancy by comparing the column densities for scattering and absorption derived for Mg and Si.  
As Fig.~\ref{f:abs} shows, for an absorbing equivalent hydrogen column density of 2.2$\times$10$^{21}$ cm$^{-2}$, the 
iron L-shell at 0.706 keV is
clearly measured. 
In the scattered spectrum iron is hardly detectable. The best fit model (Fig.~\ref{f:mrn_contr}) requires a certain amount of Fe, although not precisely quantified, 
locked in silicates, with a Mg to Fe ratio of $\sim$5:(1.4-2.5). The depletion of Mg, as well as Si, depends on the
density of the dust environment. Differently, Fe is found to be highly depleted (80-100\%) from the intercloud medium to the densest environments. Thus,
we can suppose that Mg is more abundant in silicates than Fe. In particular, a mixture of olivine and pyroxene with a ratio of 5:2 for Mg:Fe, would
totally account for the depletion of Mg and Si \citep{whi03}. In this frame, less than half of the available iron grains are locked in silicates and the
rest in other forms. The best fit of the scattered halo at $\sim$4.8$^{\prime}$, within errors, is in agreement with this simple prediction. 
If the lower limit is taken, there would be more
room for iron to be locked in other forms.    
\citet[][]{1999ApJ...517..292W} suggested, for instance, that up to 60\% of Fe could be associated with graphite to form very small grains ($a\sim 10-15$
\AA).
A Mg:Fe ratio of 1:1, as prescribed by the D03 dust mixture, fails to interpret 
the spectral energy distribution of the dust toward \cy. This again reinforces the idea that more compounds containing Mg and Si should play a role in the scattering.  
As shown in Fig.~\ref{f:mrn_contr}, the data requires a major contribution of Mg$_2$SiO$_4$ rather than iron compounds like FeSiO$_3$. 
This dust mixture leads to a stronger contribution of Mg with respect to the D03 model.\\ 
This 
interpretation is influenced by the PSF subtraction, which can artificially enhance the depth of the Si feature 
(but not significantly for Mg which lies at 1.3\,keV, where the halo is well
above the the PSF) and the continuum around it (\S~\ref{par:hspe}). 
Another unknown is the role of carbon. If the
contribution of carbon between 0.4 and 2\,keV varies substantially from the 20$-$30\% of the total budget, then the contribution of silicate would need to be revisited. 
The quality of spectral data does not allow to appreciate a substantial discrepancy between the WD01 and MRN dust size distribution. As discussed below (\S~\ref{par:d_size}), 
at the scattering angles we consider for the spectral analysis (around 4.8\pr), the bulk of the scattering is caused by grains with quite ``standard" sizes and in this case the MRN and WD01 do not differ
dramatically. However, with a deeper observation we would differentiate more between the two models at each scattering angle and this would influence the relative contribution to the scattering of the
different compounds. 
We have seen that the modeling of Si suffers from the highest uncertainty in this analysis and we cannot draw any conclusion on the basis of this feature. However, 
the prominence of the Mg feature suggests a significant contribution of magnesium compounds.    
The oxygen region of the scattered spectrum is well interpreted by D03. 
This is because of the dielectric functions, which include XAFS near the edge energy (D03). 
When convolved with the spectral resolution of these data, such sub-structures are almost
totally canceled: the difference in depth between the D03 and the \citet{1993ADNDT..54..181H} dielectric functions are indeed 
$\sim$1\% for Mg and Si, and $\sim$20\% for
iron. Oxygen
is the only element where the discrepancy is noticeable ($\sim$ a factor two).    
Evaluating the contribution of the single elements, assuming that C, O, Fe, Mg, and Si are causing the totality of the scattering, we find that oxygen can be completely
explained in terms of silicate. This supports the idea that oxygen is preferably locked in these materials, leaving little room to other O compounds 
\citep[e.g. OH, H$_2$O,][]{2001ApJ...550..793W}, at least in the diffuse ISM.
The features of the scattered halo, as the one shown in Fig.~\ref{f:data_mg2sio4}, were predicted by the scattering theory \citep[e.g.][D03]{ha70,pk96}. The 
dust physical parameters extracted from our modeling seems to bolster this interpretation. Another possibility for producing a complex halo spectrum would be an uneven
absorption at an angular scale of arcminutes around the source. If, for example, 
the absorption toward the central source is few percent less than at the outer parts, the halo/source ratio 
will show extra absorption at soft energies, similar in shape to what we observe (Fig.~\ref{f:halo_sp_allrings}).
In this case, at the energy of the absorption edges, we would expect also broad and asymmetric residuals.  
However, we found the spectral shape of the halo to have circular symmetry around the central source and this is in
conflict with a clumpy structure of the medium, and the features we observe in the scattered spectrum seem not to have an asymmetric, edge-like, shape.    
Moreover, there should be regions where, on the contrary, $N_{\rm H}$ is lower than toward the central source. 
In this case, the halo/source ratio should show an excess at soft
energy and this was not observed.    
Although the assumption that in the diffuse ISM dust is homogeneously distributed is surely simplistic, 
in the case of \cy\ we cannot prove that spatial variations of N$_{\rm H}$ on arcminutes scales play a major role in the halo shape.     


\subsection{Oxygen in the ISM}
The most prominent absorption features in the RGS band arise from oxygen transitions. The O{\sc i} resonant line corresponding to the 1s-2p
transition is clearly detected. Whether we can observe O \citep{tak03} in molecular
form is not obvious. The uncertainty is enhanced considering
the lack of laboratory measurements for oxygen locked in compounds at
these energies. The RGS resolution does not allow us to give a unique
interpretation of the spectral region at energies higher than the oxygen
edge. In terms of $\chi^2$, the addition of one or possibly two more
absorption edges or, alternatively, modeling the spectrum with three absorption lines (O{\sc i} 1s-3p, O{\sc ii} 1s-2p, O{\sc iii}, 1s-2p), provides similar results. 
From the curve of growth
analysis, \citet{juett03} calculate that the contribution of the ionized oxygen should be 10\% of
the neutral phase. 
The mildly ionized oxygen would then be produced by charge
exchange with O{\sc i}. However, in principle, absorption by oxygen in compounds is expected. 
In the present \xmm\ observation, in addition to absorption, we could also study the scattering by ID and this allows us to state that
oxygen locked in dust is actually present on the line of sight of \cy\ (Fig.~\ref{f:data_mg2sio4}).
The modeling of the halo spectrum  shows that, to first order approximation, 
the oxygen feature is well explained if all the oxygen available in dust is locked in
silicates. This translates into an oxygen to hydrogen ratio O$_{\rm dust}$/H$=180$ ppm \citep{car96} in the line of sight of \cy.   
If the oxygen abundance O/H is taken to be 490 ppm \citep{wam00}, then the depletion value for O, defined as the gas to total ISM ratio, is   
0.63. With the total hydrogen column density derived from the RGS measurement (N$_{\rm H}\sim2.2\times10^{21} {\rm cm}^{-2}$), 
we can then predict an absorption column density for oxygen locked in (silicate) dust. We obtain 
N$_{\rm abs}^{\rm O}$ for dust $\sim3.85\times10^{17} {\rm cm}^{-2}$. 
In the \citet{tak03} model (column 1, Tab.~\ref{t:rgs_bothfit}), the edge attributed to oxygen in dust has an absorption optical depth of $0.27\pm0.06$ at 0.536\,keV. 
Using the absorption cross section $\sigma^{\rm O}_{\rm abs}=1.78\times10^{-19} {\rm cm^2}$, \citep[D03, ][]{li95}, the column density for this edge is 
$1.5^{+0.3}_{-0.4} \times10^{17} {\rm cm^{-2}}$, which is too low compared to the theoretical value we obtained above.  
One possibility is that other absorption edges by oxygen in solid form are missing the detection. Indeed the energy and the structure of the edges 
slightly changes depending on which compound of oxygen and silicon is considered \citep[e.g,][]{2002RaPC...65..109O,1995PhyB..208..220W}, 
and therefore some of them may be blurred and difficult to detect with the present resolution. 
This calculation strongly depends on the adopted value of $\sigma^{\rm O}_{\rm abs}$. 
Here we considered the peak absorption cross section for oxygen locked in Mg$_2$SiO$_4$ 
\citep[D03, ][]{li95}, as determined by D03, with the underlying assumption that the oxygen edge profile is the same as the Si profile. However, small variations of 
$\sigma^{\rm O}_{\rm abs}$, lead to substantial variations of the value of N$_{\rm H}$. 
In fact, if the value $\sigma^{\rm O}_{\rm abs}$ is closer to what predicted for atomic oxygen 
($\sigma=4.8\times10^{-19}$, Verner \& Yakovlev 1995; $\sigma=3.85\times10^{-19}$, Henke, Gullikson \& Davis 1993), then the column density associated to the 0.536\,keV edge 
reconciles to the value we obtained from the scattered radiation analysis. 
As a cautionary note, the effective area of EPIC-pn diminishes toward lower energy (60\% less at 0.54\,keV than at 1\,keV) and the statistics at the oxygen 
feature may prevent us
from a full description of the dust contribution to the total extinction. 
Moreover, the value for the total O/H ratio 
is not yet solidly established \citep[][ and references therein]{jen05}. Finally, the knowledge of the theoretical 
wavelengths of these dust absorption features is very fragmentary, making their interpretation somehow aleatory.    
Taking the N$^{\rm O}_{\rm abs}$ derived starting from the scattering analysis (and $\tau={\rm N}^{\rm O}_{\rm abs}\sigma^{\rm O}_{\rm abs}=0.68$, if D03 value is taken for
$\sigma$) 
as an upper limit, 
absorption by oxygen in dust should be visible in absorption (edges and
lines) using the RGS, and
instruments with even higher energy resolution \citep{tak03}. 

\subsection{The dust size distribution}\label{par:d_size}
In this \xmm\ observation,  
the pile-up prevents us from studying the halo at
angular radii smaller than $\sim$40$^{\prime\prime}$.
This observational limit also hampers any 
detection of scattering either by large grains or standard grains located very close to the emitting source \citep{pk96}. 
Indeed these two conditions have the same effect of producing very narrow halo components in the profile.  
The data modeling of both the halo spatial profile and the halo spectral distribution show 
that both the MRN and WD01 dust size distribution can be applied despite the different amount of dust predicted for different scattering angles (Fig.~\ref{f:cy_prof}).
Grains with sizes $>0.25 \mu{\rm m}$ play a more important role in the WD01 model for radii $\ltsim 200^{\prime\prime}$ at 1 keV.  
In the size range 0.001-0.25 $\mu$m, the MRN and WD01 model do not
differ dramatically for $R_V$=3.1 \citep[see Fig.~2 of ][]{wd01} and for the grain sizes roughly between 0.1-0.2$\mu$m typically produce 
the bulk of the scattering halo at angular radii between 100$^{\prime\prime}$ and 1000$^{\prime\prime}$  \citep[e.g, ][]{1991ApJ...376..490M,drainetan}, 
which is the region that
could be directly studied in the SBP of \cy. 
The contribution of small grains ($a\ltsim 0.05$) would be best investigated if we could access the region beyond 1000$^{\prime\prime}$, but unfortunately, 
in this observation the halo begins to
fade, making the modeling challenging beyond $\sim$650$^{\prime\prime}$.\\ 
The value of the scattering optical depth derived from the SBP at 1 keV ($\tau_{\rm sca}=0.054\pm0.018$) is larger, but consistent within the errors, with the value derived by the ROSAT halo 
($\tau_{\rm sca}\sim0.039$, PS95) obtained at $\sim$1.06 keV, if the
MRN distribution is used. The PS95 modeling indeed started also from the MRN model, but leaving free some parameters in it, among 
which the maximum grain size and
the slope of $n(a)$. Their best fit requires a maximum grain size of only 0.15 $\mu$m. This makes the halo model flatter and the derived value of $\tau_{\rm sca}$ lower than
what we measure. 
The WD01 distribution predicts instead almost the double ($\tau_{\rm sca}=0.067\pm0.018$) of the ROSAT result.  
This trend is visible at all energies (Fig.~\ref{f:tot_tau}). This is mostly due to the increased scattering
``power" by
larger grains at smaller radii predicted by the WD01 (Fig.~\ref{f:cy_prof}).\\ 
When the total scattering optical depth (for both MRN and WD01 case) as a function of energy is compared with a theoretical model (D03, Fig.~\ref{f:tot_tau}), 
we see that it does not measure a substantial
part of the scattered radiation. Indeed the model over-predicts the data at all energies apart from perhaps the last point. 
The distribution of $\tau_{\rm sca}$ flattens toward lower energies, while the model, 
is significantly steeper. As noted in D03, the discrepancy was also found for other halo analysis 
\citep[e.g., PS95, ][]{gx13,1994ApJ...436L...5W}. 
The \citet{drainetan} measurement instead 
has been found
in agreement with the model (D03).\\ 
The value of $I_{\rm frac}$ (and thus of $\tau_{\rm sca}$), extracted at each energy, is model dependent and, moreover, 
can be substantially influenced by instrumental effects. 
For example, the scattering angle range accessible to this study ($200^{\prime\prime}-600^{\prime\prime}$) strongly privilege the observation of the scattering of  
$1-2$ keV photons \citep[e.g., ][ D03]{1991ApJ...376..490M}, while the bulk of the emission for softer photons peaks at larger scattering angles.     
The estimated values for the total scattering optical depth, and in particular the ones related to the soft scattered photons are certainly lower limits. 
A significant fraction of the halo, in the form of fainter or narrower components, may be masked by the PSF wings (at small radii) or unaccessible
because of the faint scattered emission with respect to the background (at larger radii).

\subsection{The dust distribution along the line of sight}
The distribution of dust along the line of sight could be studied through the SBP of the halo. The distribution of dust seems to be evenly distributed, 
at least for a fractional distance of the total path $x<0.6$, corresponding to a linear distance between 4.3-6.7 kpc, \citep[depending on the distance
estimates for \cy; ][]{1999MNRAS.305..132O,1998ApJ...498L.141S}.   
However, if the dust distribution is imposed to be uniform up to a fractional distance of 0.99 the fit worsens significantly suggesting that our line of sight passes
through different dust clumps.  
A contribution from scattering events closer to the source is indeed likely to be present, but fainter or narrower halo components, as detected in other sources 
\citep[e.g.][]{gx13,drainetan,cos04}, cannot be investigated using \xmm\ for radii $<12^{\prime\prime}$ (and in any case not for \cy\ because of the pile-up).  

\section{Conclusions}
 
We have presented \xmm\ results on the effect of scattering and absorption by ID along the line of sight to the bright X-ray binary \cy . 
This study led
to the unprecedented detection of the 
elements in the ID responsible for the X-ray scattering: oxygen, magnesium, and silicon. 
To first order, the modeling of the pure scattered radiation suggests a major
contribution of silicates in the form of olivine and pyroxene, in the energy range 0.4-2 keV. 
The best fit of the scattered spectrum shows that the ratio of Mg to Fe, locked in dust grains, is 5:(1.4:2.5). This is consistent with
a picture in which Mg and Si are for the most part locked in silicates \citep{whi03}.\\
The contribution of carbon, a fundamental constituent of ID, could not be quantified as its most prominent feature (0.28\,keV) 
lies below our selected EPIC-pn energy band.  
In the RGS spectrum, we studied the complexity around the oxygen edge, investigating the possibility of absorption 
by atomic and molecular oxygen, 
as suggested by Takei
et al (2002), in comparison with absorption by atomic neutral and ionized oxygen \citep{juett03}. 
The RGS resolution is not sufficient to prefer one interpretation to
the other, but the complementary information from the EPIC-pn analysis of the scattered spectrum allows
us to detect oxygen locked in dust, preferably in the form of silicates. 
Starting from the scattered halo spectral modeling, we estimated that the absorption column density we expect by oxygen locked in silicates is indeed 
measurable using the RGS. The value of this column density seems too high compared 
to what measured for the absorption edge that \citet{tak03} interpret as arising from dust. However,
instrumental/theoretical uncertainties makes this result not conclusive.\\
The study of the spectral energy distribution of the scattered radiation, performed in the halo region where the signal-to-noise ratio was best (around
$\sim$4.8$^{\prime}$), stressed the need of using an accurate theoretical approach to the data.
The full Mie theory had to be used to model the data satisfactorily,
especially below 2 keV where the chemistry of the halo can now be
studied. This approach was already applied to ROSAT data \citep{sd98,drainetan}.
With \xmm\ we could extend this analysis, performing spatially resolved spectroscopy of the halo that could not be interpreted unless the Mie
differential scattering cross section was used.\\
The modeling of the SBP shows that the dust is uniformly distributed along the line of 
sight at least for a fractional distance of the total
path $x<0.6$, corresponding to a linear distance between 4.3-6.7 kpc, depending on the source distance estimates. 
However, a uniform dust distribution along the complete path toward \cy\ is not required by the data, 
hinting to a clumped structure of the dust for $x>0.6$. 
Within the instrumental uncertainties, the data are acceptably fit by both a MRN and a WD01 dust size distribution. 
The inferred scattering optical depth is approximately 0.054 and 0.067 at 1 keV for the MRN and WD01 distribution, respectively. 
We extended the modeling of the SBP of the
halo to the 0.4-2 keV band. The derived values of the total $\tau_{\rm sca}$ as a function of energy 
are systematically lower than what predicted by the theory (D03), pointing 
out that some halo components may be easily missed due to instrumental effects.

\begin{acknowledgements}

The authors thank the referee, Prof.~B.T.~Draine, for his valuable comments which improved the quality of the paper. 
This project was developed for the most part during the stay of E.C. at    
the Max-Planck-Institut f\"ur extraterrestrische Physik. 
E.C. wishes to thank also K.C.~Steenbrugge and F.~Verbunt 
for carefully reading
this manuscript.
The XMM-{\em Newton} project is supported by the Bundesministerium f\"ur
Bildung und Forschung / Deutsches Zentrum f\"ur Luft- und Raumfahrt 
(BMBF/DLR), the Max-Planck-Gesellschaft and the Heidenhain-Stiftung.

\end{acknowledgements}

\end{document}